\documentclass[aps,prb,reprint,groupedaddress]{revtex4-1}

\usepackage{amsmath,amssymb} 
\usepackage{amsmath} 
\usepackage{amsfonts} 
\usepackage{bm} 
\usepackage[colorlinks=true,citecolor=blue,linkcolor=blue]{hyperref}
\usepackage[pdftex]{graphicx} 
\usepackage{braket} 

\begin{document}

\title{Phase diagram of the quantum spin-1/2 Heisenberg-$\Gamma$ model on a frustrated zigzag chain}

\author{Hidehiro Saito}
\email{saito-hidehiro722@g.ecc.u-tokyo.ac.jp}
\author{Chisa Hotta}
\affiliation{Department of Basic Science, University of Tokyo, Meguro-ku, Tokyo 153-8902, Japan}
\date{\today}

\begin{abstract}
We investigate the quantum spin-1/2 zigzag chain with frustrated $J_1$-$J_2$ Heisenberg interactions, 
incorporating additional off-diagonal exchange interactions known as the $\Gamma$ term, 
both with and without an applied magnetic field. 
Based on the density-matrix renormalization group calculation, 
we map out the ground state phase diagram that shows a variety of magnetic and nonmagnetic phases 
including multicritical points and several exactly solvable points. 
Upon introducing a finite $\Gamma$ term, we observe the persistent dimer singlet state 
of the $J_1$-$J_2$ Heisenberg model, sustaining a nonzero spin gap, 
while also giving rise to a gapless nonmagnetic excitation, 
manifesting in the substantial zero-energy peak in the nematic dynamical structure factor. 
This gapless peak-mode remaining almost as a fluctuation to the ground state, 
induces dilute but robust concentration of nematicity on top of singlets on dimers, 
which we call the nematic singlet-dimer phase. 
When the whole nematic excited mode condenses and replaces the singlet, 
the nematic-dimer phase transforms to the Ising-type ferromagnetic or antiferromagnetic long-range orders 
that arise from the $\Gamma$ term spontaneously selecting magnetic easy axes. 
Its orientations dictate the type of magnetic order under geometric frustration effects 
as predicted by Landau's mean-field theory. 
These theoretical findings provide insights into the exotic low-temperature phase observed in YbCuS$_2$, 
characterized by gapless excitations and seemingly nonmagnetic behavior accompanied by incommensurate correlations. 
\end{abstract}

\maketitle

\section{Introduction}
\label{sec:introduction}
Exploring interesting quantum-disordered phases in materials stands by now as a major challenge in condensed matter physics. 
Longstanding intense investigations into triangular and kagome quantum spin liquids\cite{Anderson1973,Anderson1987,Zhou2017} were recently 
spurred by the discovery of Kitaev spin liquids featuring Majorana quasiparticle excitations\cite{Kitaev2006}, 
and it motivated the detailed examination of new types of quantum anisotropic exchange interactions 
in 4d, 5d, and 4f insulating magnets\cite{Jackeli2009,Chaloupka2010,Rau2018,Rau2016}. 
These interactions are influenced by strong spin-orbit coupling, moderate crystal field effects, and electron correlations, 
which play a crucial role in the emergence of Kitaev and $\Gamma$ terms 
alongside the previously studied Dzyaloshinskii-Moriya and ring exchange interactions. 
\par
Spin liquids are, however, not the sole focus of nonmagnetic disordered phases in quantum magnets.
There are valence-bond solids in the spin-1 chain known as symmetry-protected topological phase\cite{Pollmann2010}, 
valence-bond crystals based on singlets in Shastley-Sutherland model\cite{Miyahara1999,Nomura2023}, 
and spin nematic phases or a quadrupolar order triggered by the condensation of two-magnon bound state 
for $S=1$ models\cite{Penc2011,Chubukov1991,Harada2002} and $S=1/2$ or $S=1$ spin ladders\cite{Hikihara2008,Lauchili2006}. 
Notably, some of these phases break lattice symmetry while suppressing magnetic orderings, making them experimentally more accessible compared to spin liquids. 
Fortunately, there exist material platforms that host these phases, 
such as $\kappa$-ET$_2$Cu$_2$(CN)$_3$\cite{Shimizu2003} and ZnCu$_3$(OH)$_6$Cl$_2$\cite{Mendels2015,Norman2016}, for spin liquids, 
NENP for Haldane chain\cite{Lu1991}, and SrCu$_2$(BO$_3$)$_2$\cite{Kageyama1999} for orthogonal dimer phases, 
providing crucial insights into the nature of these intriguing phases governed by strong quantum fluctuations and correlations. 
\par
Despite these advancements, the role of off-diagonal symmetric $\Gamma$ terms, 
which have recently been observed in Yb and other 4f-based magnets, 
remains largely unexplored \cite{Rau2018,Rau2016}. 
Initially discussed as secondary terms in Kitaev magnets, the $\Gamma$ term, 
in conjunction with the Heisenberg interaction, destabilizes Kitaev spin liquids in two dimensions\cite{Rau2014-kitaev}. 
In the one-dimensional analog known as the Kitaev-Heisenberg-$\Gamma$ chain \cite{Yang2020}, 
the $\Gamma$ term significantly alters the ground state phase diagram, 
leading to SU(2) symmetric points and magnetic orderings with spontaneously oriented easy axes.
\par
This paper elucidates the role of the $\Gamma$ term in geometrically frustrated zigzag Heisenberg spin-1/2 chain, 
whose potential platform is the 4f insulating magnet, YbCuS$_2$. 
Previously, we have microscopically derived the quantum spin model for this material based on the $\Gamma_6$ Kramers doublet of Yb ions that forms a zigzag chain, 
revealing nearly isotropic $J_1\sim J_2$ Heisenberg interactions and small but finite $\Gamma$-type exchange couplings \cite{Saito2024}. 

Experimentally, YbCuS$_2$ undergoes a first-order transition to a low-temperature phase lacking clear long-range magnetic ordering, 
with NMR suggesting gapless nonmagnetic excitations \cite{Ohmagari2020,Hori2023}. 
The experimental magnetic-field-temperature phase diagram does not conform to the previous theory of the simple $J_1$-$J_2$ Heisenberg model\cite{Hikihara2010}. 
Our theoretical parameterization considers the $J_1$-$J_2$ and $\Gamma_1$-$\Gamma_2$ zigzag chain, 
and unveil the entire phase diagram with and without a magnetic field, incorporating ferromagnetic, antiferromagnetic, or mixed Heisenberg exchange couplings. 
We employ techniques such as density matrix renormalization group (DMRG), exact diagonalization, 
bond-operator approach, mean-field analysis on Ising competing orders. 
The key finding is the emergent nonmagnetic gapless excitations introduced at an infinitesimally small value of $\Gamma$ 
inside the robust magnetic spin gap of the dimer singlet long-range ordered phase. 
On top of that, the competing various magnetically and nonmagnetically ordered phases 
appear as the complex interplay of geometrical frustration effect of zigzag structure 
and the competition of Heisenberg and $\Gamma$-terms. 
\par
The paper is structured as follows: 
Section \ref{sec:model} presents the model Hamiltonian and the ground-state phase diagram in detail, 
Section \ref{sec:bond} focuses on antiferromagnetic cases and the effect of the $\Gamma$ term using the bond-operator approach 
and mean-field analysis, Section \ref{sec:mag} outlines the magnetic phase diagram, 
and Section \ref{sec:summary} discusses materials and experimental implications.

\section{model and the ground state phase diagram}
\label{sec:model}
\begin{figure*}
    \centering
    \includegraphics[width=18cm]{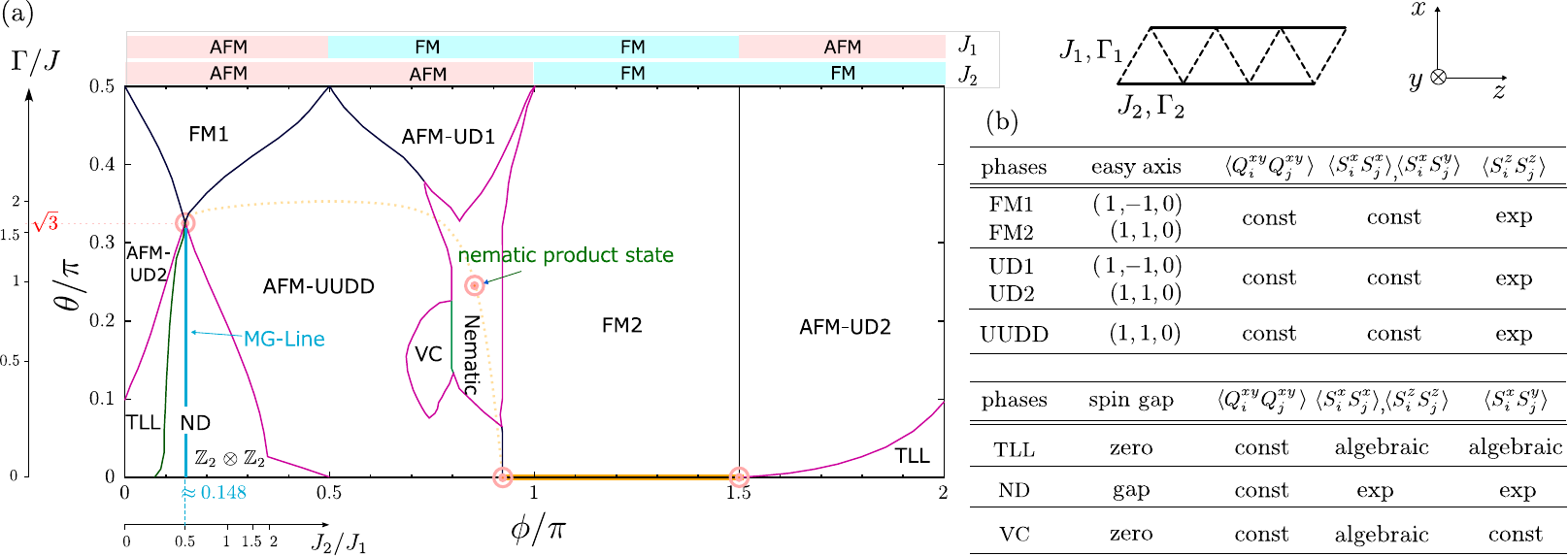}
    \caption{(a) Ground-state phase diagram of a spin-1/2 zigzag spin chain shown in the right panel, 
     where the $z$-axis parallel to the legs and the $y$-axis perpendicular to the triangular plane. 
     We take $J_1=\cos\phi\cos\theta, J_2=\sin\phi\cos\theta,\Gamma_1=\cos\phi\sin\theta$ and $\Gamma_2=\sin\phi\sin\theta$. 
     Three nonmagnetic phases are denoted as the nematic-singlet dimer (ND) phase, the Tomonaga-Luttinger liquid (TLL) phase, 
     and vector chiral (VC) phase. 
     Magnetically long-range ordered phases are ferromagnetic (FM), antiferromagnetic (AFM) UD, or UUDD phases. 
     Four circles represent the exact solution points 
     (tricritical Lifshitz point, nematic product state, RVB state, fully polarized ferromagnet), 
     and the bold lines (Majumdar-Ghosh line and $\theta=0$ line) are the exact solution lines. 
     Pink and dark phase boundaries are determined by the change of magnetization and the energy crossing, respectively, 
     and the VC and TLL-ND phases are from the scaling and vector chiral correlations, respectively. 
     (b) Characteristics of phases in terms of magnetic order, correlation functions, and spin gap, 
      to be detailed in Figs.~\ref{f2}-\ref{f5}.
}
    \label{f1}
\end{figure*}
\subsection{Model Hamiltonian}
We consider a quantum spin-1/2 Hamltonian on a zigzag chain given as
\begin{equation}
    {\cal H}= \sum_{j} \sum_{\eta=1,2} 
    J_\eta \bm S_j \cdot \bm S_{j+\eta} 
    + \Gamma_\eta (S^x_jS^y_{j+\eta}+ S^y_{j}S^x_{j+\eta}), 
     \label{eq:ham}
\end{equation}
where $J_\eta$ and $\Gamma_\eta$ are the Heisenberg and anisotropic exchange interactions between 
nearest ($\eta=1$) and next nearest ($\eta=2$) spins. 
We consider both the antiferromagnetic(AFM) and ferromagnetic(FM) couplings of $J_\eta$ and set  
\begin{align}
&J_1=\cos\phi\cos\theta,\quad J_2=\sin\phi\cos\theta, \notag\\ 
& \Gamma_1=\cos\phi\sin\theta,\quad \Gamma_2=\sin\phi\sin\theta, 
\end{align} 
where dividing the parameter range into four, 
$\phi=[0:\pi/2], [\pi/2:\pi], [\pi:3\pi/2], [3\pi/2:2\pi]$ correspond to AFM-AFM, FM-AFM, FM-FM, AFM-FM interactions of $J_1$-$J_2$, respectively. 
The sign of $\Gamma_{\eta}$ can be converted by the local unitary transformation and does not influence the physical state. 
The spin quantization axis $z$ is taken parallel to the chain (see Fig.~\ref{f1}). 
\par
The model at $\Gamma_{\eta}=0$ corresponds to the zigzag Heisenberg spin chain, 
the phase diagram of which has been extensively studied previously. 
In the AFM-AFM case where $J_1$ and $J_2$ are both positive, 
there exists a transition from a Tomonaga-Luttinger liquid (TLL) phase to a dimer singlet phase characterized 
by a finite spin gap \cite{Majumdar1969,Majumdar1969-2,Haldane1982}. 
This transition occurs at a critical point $(J_2/J_1)_c\approx0.2411$ ($\phi\approx 0.075\pi$) \cite{Okamoto1992,Eggert1996}. 
Within the dimer singlet phase, a Lifshitz point $(J_2/J_1)_L\approx0.5206$ ($\phi\approx 0.153\pi$) delineates two distinct regions: 
one with commensurate $q=\pi$ short-range magnetic correlations and another with incommensurate $q<\pi$ correlations 
at lower and higher $J_2/J_1$ values, respectively \cite{Hikihara2001,Bursill1995}. 
There is also an ongoing debate regarding a potential transition from the dimer phase to a gapless phase 
around $J_2/J_1\approx 2.2$  ($\phi\approx 0.36\pi$), identified through level crossing experiments \cite{Kumar2015}, 
exact diagonalization and density matrix renormalization group (DMRG) studies \cite{Soos2016}. 
This observation contradicts field theory predictions, which suggest a finite gap $\Delta \sim \exp (-(J_2/J_1)^{\eta})$ with $\eta=1$ \cite{White1996} or $\eta=2/3$ \cite{Itoi2001} for large $J_2/J_1$ values. 
We see shortly in our phase diagram in Fig.~\ref{f1}(a) that 
the phase boundary has a kink which extrapolates to $J_2/J_1\approx 2.2$, 
indicating that the phase changes its nature at $J_2/J_1\gtrsim 2.2$. 
Whereas, it is numerically difficult to conclude whether 
there is a finite but exponentially small spin gap or not. 
\par
In the FM-AFM case, a Haldane dimer phase exists where ferromagnetic spin pairs form $S=1$ states for $J_2/J_1<-1/4$ ($0.922\pi<\phi<\pi$) \cite{Furukawa2012,Agrapidis2019}. 
At the critical point $J_2/J_1=-1/4$ ($\phi=0.922\pi$), an exactly solvable resonating valence bond (RVB) state appears \cite{Hamada1988}, 
leading to a transition to a phase with ferromagnetic (FM) long-range order. 
Similar to the AFM-AFM case, the existence of a gap at large $J_2/|J_1|$ remains a subject of controversy \cite{Kumar2015,Agrapidis2019}. 
\par
For the FM-FM and AFM-FM regions, there is no frustration and the ferromagnetic and TLL phases are realized, respectively. 
\par
The zigzag XXZ chain exhibits a richer variety of phases because of the broken SU(2) symmetry, 
including dimer singlet, spin fluid, ferromagnetic, antiferromagnetic phases\cite{Somma2001,Plekhanov2010,Hirata2000,Gerhardt1998,Jafari2007}, 
and long-range order of vector chirality at large Ising anisotropy\cite{Hikihara2001}. 
Similarly, a zigzag Heisenberg ladder in an applied magnetic field also displays comparably 
a Tomonaga-Luttinger liquid (TLL), a $1/3$-magnetization plateau, spin-density-wave, vector chiral, 
and fully polarized phases \cite{Okunishi2003,Okunishi2008,Hikihara2010,Hikihara2008,Sudan2009}.
\par
Here, we investigate the impact of the anisotropic exchange interaction known as the $\Gamma$ term on these previously studied phases.
\subsection{Ground-state phase diagram}
\label{sec:ground}
\subsubsection{Overview of the phase diagram}
Figure~\ref{f1} shows the phase diagram on the plane of 
$\phi=\arctan(J_2/J_1)$ and $\theta=\arctan(\Gamma_{\eta}/J_{\eta})$. 
Among four different regions separated by the sign of exchange interactions 
(see the sign of $J_1$ and $J_2$ in the top panel), 
we have previously studied the AFM-AFM region ($\phi\le 0.5$) in Ref.[\onlinecite{Saito2024-2}] 
with a particular focus on the multicritical point observed at $J_2/J_1=1/2, \Gamma/J=\sqrt{3}$ 
$(\phi\approx0.148\pi,\;\theta=\pi/3)$. 
This point is rigid as we find an exact solution that has degeneracy of order-$N^2$ 
as shown in Ref.[\onlinecite{Saito2024-2}]. 
It is not only a multicritical but also a Lifshitz point, 
because the uniform ferromagnetic (FM1) and the antiferromagnet with up-up-down-down period (AFM-UUDD) 
meet the quantum disordered two phases, i.e. the TLL and nematic-singlet dimer phases. 
The low energy effective theory at around this multicritical point is shown in Appendix~\ref{app:mean}, 
which nicely explains the Ising types of competitions among AFM-UD2 (antiferromagnet with up-down period), 
UUDD, and FM1 phases, which are summarized in Fig.~\ref{f1}(b). 
In the following subsections, we study the details of these phases. 
\par
There are three other exact solutions (see Appendix~\ref{app:exact}), 
the nematic product state at 
$(\phi/\pi,\theta/\pi)=(0.852,0.25)$ ($J_2/J_1=-1/2,\Gamma/J=1$), 
the RVB solution\cite{Hamada1988}, and decoupled ferromagnetic chain, 
which are linked by bold lines and also partially host exact solutions (for nonbroken lines). 
These states are obtained using the method we developed 
to have the exact MPS-based solutions for frustration-free models\cite{Saito2024-3}. 
\par
Here, we briefly explain how we identify the nematic product exact ground state. 
At $\;\Gamma_{\eta}=J_\eta$ and $J_2/J_1=-1/2$, 
the Hamiltonian has a typical frustration-free form, 
${\cal H}=\sum_{l} \hat h_l$, 
given as the sum of operator $\hat h_l$ acting on the $l$-th triangle, 
\begin{align}
\hat h_l= \sum_{i,j\in l} (-)^{|i-j|} J(\bm S_{i} \cdot \bm S_{j} + S^x_{i}S^y_{j} + S^y_{i}S^x_{j}),
\end{align}
where we set one AFM and two FM bonds with $J\equiv J_2$. 
In a triangular unit, $\hat h_l$ has four-fold degenerate ground states of energy $-3J/4$ 
and the four excited states with energies $\pm 2(\sqrt{3}+3)J/4$. 
Using 0/1 representing up/down spins on $[l+1, l, l-1]$ sites on the triangle in the descending order,
they are given as 
\begin{align}
& |000\rangle+i|011\rangle=|0\rangle \otimes (|00\rangle + i|11\rangle), \notag\\ 
& |000\rangle+i|110\rangle=(|00\rangle + i|11\rangle)\otimes |0\rangle , 
\label{eq:nemaexact}
\end{align}
and their time reversal states, $|111\rangle-i|100\rangle$, and $|111\rangle-i|001\rangle$. 
The energy of the Hamiltonian consisting of $N$ triangles is $-3JN/4$ at the lowest, 
and the product state $|\Psi\rangle=\prod_{i=1}^{N/2}|p_1\rangle_{2i-1,2i}$, 
with $|p_1\rangle=(|00\rangle+i|11\rangle)/\sqrt{2}$, satisfies such energy condition, 
because for all choices of triangles, we find either of Eq.(\ref{eq:nemaexact}) to be the constituent. 
Its time reversal or translational counterparts are the other degenerate ground state. 
The translational symmetry is broken similarly to the Majumdar-Ghosh (MG) singlet product state at $J_2/J_1=0.5$. 
The other exact solutions in the phase diagram can be obtained in the same frustration-free form, 
although the other non-product state solutions are not explicitly written in the analytical form 
but rely on the MPS language\cite{Saito2024-3}. 
\par
The phase diagram was determined using the DMRG method\cite{White1992,White1993} and the exact diagonalization (ED) method. 
For DMRG we calculate the system typically of size $N=100$ and keep up to $\chi=200$ states with up to 160 sweeps. 
Because the model with quantum anisotropy often exhibits orders or 
correlations with incommensurate or unexpected periods, 
we adopt the sine-square deformation (SSD)\cite{Nishino2011} 
that suppresses the numerical biases often induced in finite-size clusters; 
for any given Hamiltonian ${\cal H}=\sum_j \hat h(r_j)$ based on the local operator $\hat h(r_j)$ 
at spatial coordinate $r_j=1,\cdots, N$, 
the method deforms its local energy scale by the envelope function as 
${\cal H}_{\rm ssd}=\sum_j \hat h(r_j) f(r_j)$, using the $f(r_j)=\sin^2(\pi r_j/(N+1))$. 
This sine-square function $f(r_j)$ takes a maximum at the center of the system and goes to zero at both edges. 
It is proved both numerically and analytically that the SSD Hamiltonian offers a quantum ground state 
equivalent to that of the periodic boundary condition (PBC) \cite{Hikihara2011,Maruyama2011,Katsura2011}. 
Additionally, it has two advantages: 
the damping of the finite size effect and the ability to capture incommensurate orders very accurately 
by avoiding the bias to wavevectors commensurate with the system size\cite{Hotta2012,Hotta2013}. 
The boundary effects are safely excluded and the correlation functions as well as local quantities 
are safely evaluated, much more reliably than the open boundary ones\cite{Shibata2011}. 
The DMRG using SSD also can evaluate the continuous magnetization curve very accurately, 
which is particularly useful here because the model does not conserve the total magnetization 
and the standard evaluation of the spin gap as the difference in the energy of total $S=0$ and $S=1$ states is not available. 

\subsubsection{Phase boundaries}
\label{sec:phaseboundary}
We first introduce the order parameters of the three magnetic phases and the two nonmagnetic phases in the diagram: 
\begin{align}
\text{FM}\rule{10mm}{0mm} &: S^\alpha_i+S^\alpha_{i+1}+S^\alpha_{i+2}+S^\alpha_{i+3}, \notag\\
\text{AFM-UD}\rule{5mm}{0mm} &: S^\alpha_i-S^\alpha_{i+1}+S^\alpha_{i+2}-S^\alpha_{i+3}, \notag\\
\text{AFM-UUDD} &:  S^\alpha_i\pm S^\alpha_{i+1}-S^\alpha_{i+2}\mp S^\alpha_{i+3}, \notag\\
\text{VC}\rule{10mm}{0mm} &:  \kappa_i^{(n)}=S_i^x S_{i+n}^y-S_i^y S_{i+n}^x, \notag\\
\text{N}\rule{12mm}{0mm} &: Q^{xy}_i=S_i^x S_{i+1}^y+S_i^y S_{i+1}^x. 
\label{eq:orderpm}
\end{align} 
For the VC phase, we choose $\kappa_i^{(1)}$ which is the most susceptible to the present Hamiltonian. 
Regarding the nematic (N) phase, 
There are five independent nematic (quadrupolar) operators 
defined on a spin-1/2 pairs forming spin-1 given as \cite{Andreev1984}
\begin{align}
    Q_{i}^{\alpha\beta} 
    = S_{i}^{\alpha} S_{i+1}^{\beta} + S_{i}^{\beta} S_{i+1}^{\alpha}
      - \dfrac{2}{3} \left( \boldsymbol{S}_{i} \cdot \boldsymbol{S}_{i+1} \right) \delta_{\alpha\beta}.
\end{align} 
In Eq.(\ref{eq:ham}), the parameter $\Gamma$ is coupled to the $xy$ component
$Q_i^{xy}=S_i^xS_{i+1}^y+S_i^yS_{i+1}^x$, 
and it works as a ``field" (or a chemical potential) to condense the nematic particle represented by $Q_i^{xy}$, 
and accordingly, the other four parameters are irrelevant.  

\begin{figure}
    \centering
    \includegraphics[width=7.5cm]{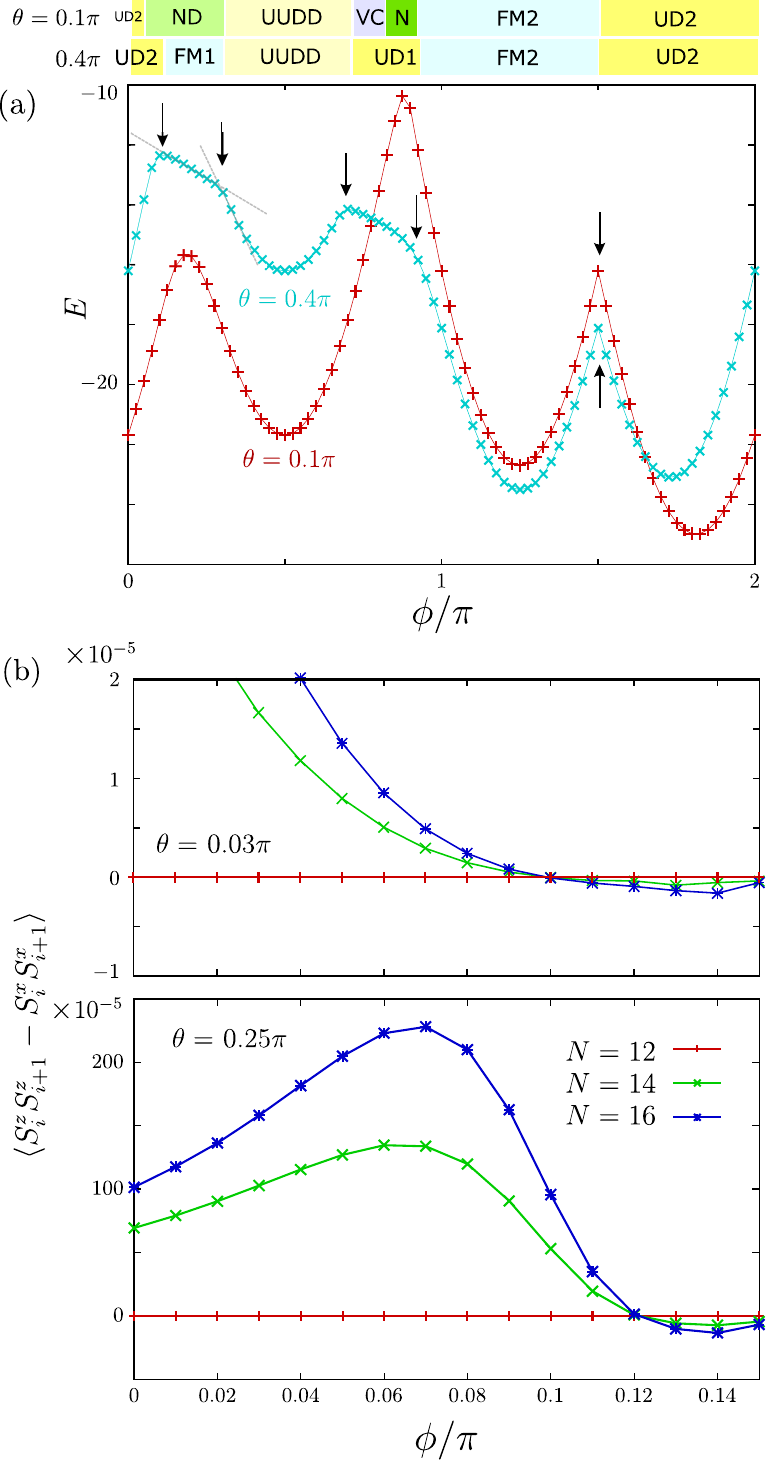}
    \caption{(a) Energy obtained by DMRG calculation and (b) $\langle S^z_i S^z_{i+1}-S^x_i S^x_{i+1}\rangle$ 
    obtained by the exact diagonalization with PBC as functions of $\phi$. 
    The kinks in (a) and crossings in (b) are used to determine the magnetic and nonmagnetic phase boundaries, respectively.
    Broken lines in panel (a) are shown as a guide to determine one of the phase boundaries. 
}
    \label{f2}
\end{figure}
\begin{figure}
    \centering
    \includegraphics[width=8cm]{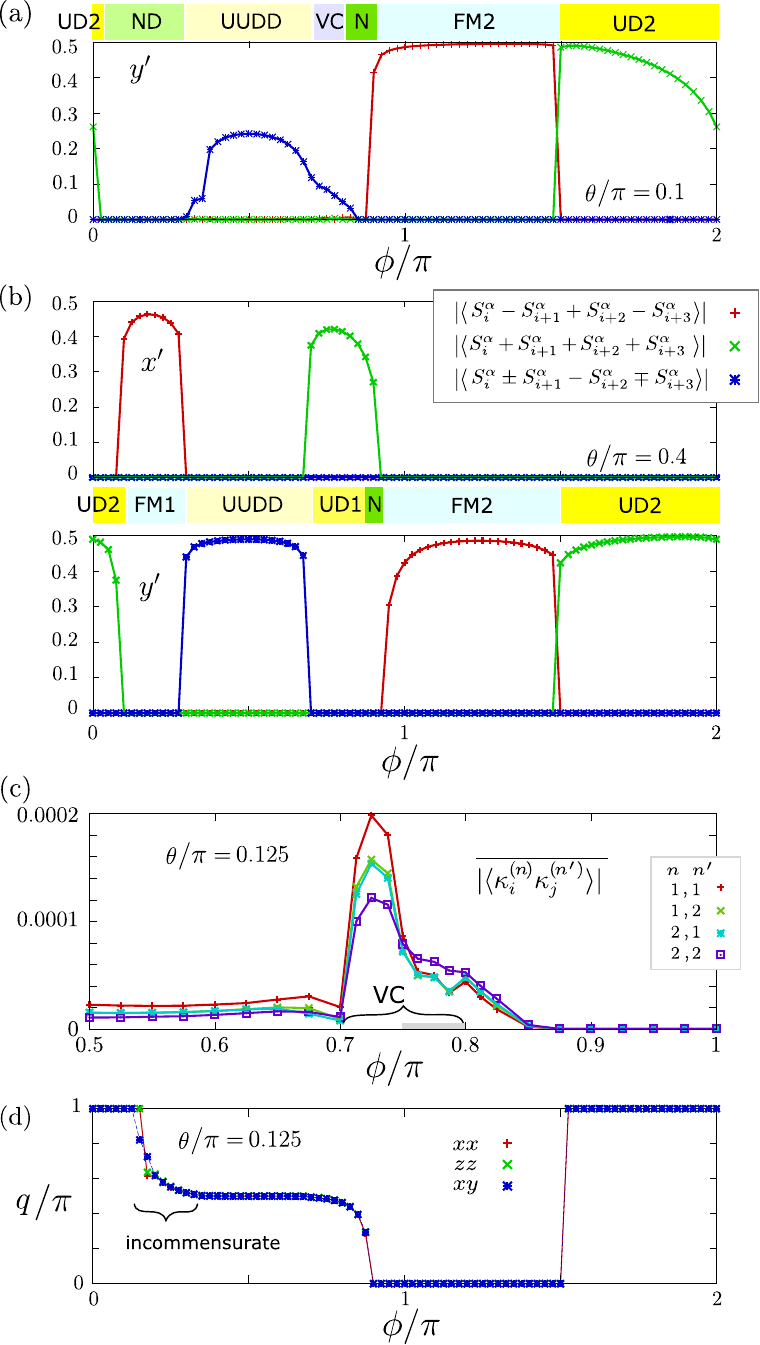}
    \caption{Uniform and staggered magnetizations in a period of two or four sites, 
    $S^\alpha_i+S^\alpha_{i+1}+S^\alpha_{i+2}+S^\alpha_{i+3}$, 
    $S^\alpha_i-S^\alpha_{i+1}+S^\alpha_{i+2}-S^\alpha_{i+3}$, and 
    $S^\alpha_i\pm S^\alpha_{i+1}-S^\alpha_{i+2}\mp S^\alpha_{i+3}$, 
    for (a) $\theta/\pi=0.1$ ($\Gamma/J\approx0.32$) in $\alpha=y'$-direction and (b) $\theta/\pi=0.4$ ($\Gamma/J\approx3.1$) in $x'$ and $y$'-directions 
    as functions of $\phi$. 
   (c) Vector chiral correlation function $\overline{|\langle \kappa^{(n)}_i \kappa^{(n')}_j \rangle|}$ 
     averaged over $|i-j|=58-61$ as function $\phi$. The gray region has the relatively weak signal of chiral correlation. 
   (d) Peak position $q$ of the structure factor ${\cal S}^{\alpha\alpha'}(q)$ for $\alpha\alpha'=xx,zz,xy$ 
    as a function of $\phi$. 
    We set $\theta/\pi=0.125$ ($\Gamma/J\approx0.41$) for (c,d). 
 }
    \label{f3}
\end{figure}
\par
The phase boundaries are determined numerically. 
The first-order transition takes place between magnetic orders of different periods and 
the second-order transitions are mostly the nonmagnetic-magnetic ones. 
We show in Fig.~\ref{f2}(a) the energy of the SSD Hamiltonian of the DMRG calculation for $N=100$ as a function of $\phi$. 
The observed kinks or the changes in the functional form provide the magnetic phase boundaries 
very accurately, 
as has been demonsrated in previous literatures\cite{Kawano2023,Makuta2024}. 
Namely, even if the envelope function $f(r_j)$ is placed, which will alter the value of the total energy of the system itself, 
because the energy is an extensive quantity, the comparison of energies between different phases works out 
quantitatively as accurate as we do in PBC. 
We can see shortly that they coincide with the ones we derive using the magnetic order parameters.  
The boundary between the nematic-singlet and the TLL phases is difficult to detect in standard methods. 
In Fig.~\ref{f2}(b) we show the $\langle S_{i}^zS_{i+1}^z-S_{i}^xS_{i+1}^x\rangle$ obtained by the 
Lanczos exact diagonalization method, which measures the anisotropy of nearest neighbor spin coupling. 
We find that the results for different $N$'s cross at the single point, 
which offers accurate scale-free boundary points separating TLL-ND phases. 
\par
The magnetic phase boundaries are clearly detected by the magnetization 
of several lattice periods and magnetization axes. 
Because we use the SSD, the values of magnetization measured at the center of the system 
is not influenced by the boundary effect and is free of finite-size effects. 
Here, we measure the magnetization along the $x'$ and $y'$ axes which are obtained by 
rotating the $x$ and $y$ axes by $\pi/4$ about the $z$-axis. 
Indeed, our Hamiltonian Eq.(\ref{eq:ham}) remains unchanged under the 
$\pi$-rotation about both the $\bm x'=(-1,1,0)$ and $\bm y'=(1,1,0)$ axes, 
where the spins are transformed as $(S^x,S^y,S^z)\rightarrow (-S^y,-S^x,-S^z)$ and $(S^y,S^x,-S^z)$, respectively. 
It is thus natural to consider the two as magnetic easy axes. 
We can thus introduce the order parameters of the magnetic phases as in Eq.(\ref{eq:orderpm}). 
In Fig.~\ref{f3}, we show these uniform and staggered magnetization along $\alpha=x',y'$ 
for two parameters across the phase diagram. 
In the case of $\theta=0.1\pi$ ($\Gamma/J\approx0.32$), $x'$-component of these magnetizations are exactly zero throughout $\phi$ and are not shown. 
Otherwise, the magnetic phases have one of the magnetizations being finite, and are exclusive to each other, 
capturing the phase boundaries very well. 
The types of magnetizations are summarized in Fig.~\ref{f1}(b). 
\par
In Fig.~\ref{f3}(d) we show the peak position of the structure factor of the 
two-point spin-spin correlation functions 
\begin{align}
{\cal S}^{\alpha\alpha'}(q)= \frac{1}{N-1}\sum_{|j-j'|=1}^{N-1} e^{iq(j-j')} \langle S_j^\alpha S_{j'}^{\alpha'}\rangle,
\end{align} 
where in the calculation we performed the SSD Fourier transformation using the envelope function \cite{Kawano2022}. 
Previous studies for the $\Gamma=0$ zigzag model reported the transition from 
a $q=\pi$ to $q<\pi$ state of the diagonal $\alpha=\alpha'$ structure factor 
inside the singlet dimer phase \cite{Hikihara2001,Bursill1995}. 
We indeed find such a transition for $\theta>0$ ($\Gamma \ne 0$) in the present model. 
In particular, in the ND phase at $0.15\lesssim \phi/\pi \lesssim 0.4$ ($0.5\lesssim J_2/J_1 \lesssim 3$), the incommensurate $q\gtrsim \pi/2$ is 
observed which transforms to the UUDD phase of period $q=\pi/2$ at larger $\phi$. 
The difference induced by the $\Gamma\ne 0$ is that 
the off-diagonal $ xy$ component appears comparable to the diagonal ones. 
\par
Let us briefly discuss the types of symmetry breakings. 
As mentioned above, the spins have $x'$ and $y'$ as magnetic easy axis, namely the 
${\mathbb Z}_2 \otimes {\mathbb Z}_2$ symmetry is preseent in the Hamiltonian. 
In the ND and TLL phases, the spins are not ordered, namely this ${\mathbb Z}_2 \otimes {\mathbb Z}_2$ is not broken. 
The ND phase breaks the translational symmetry but TLL does not. 
When the magnetic orderings take place, they break one of the ${\mathbb Z}_2 \otimes {\mathbb Z}_2$ 
(while part of ${\mathbb Z}_2$ combined with lattice translation remains), 
and the spins are polarized in one of the two easy axes. 
The translational symmetry is broken for the AFM phases, while kept for FM phase. 
The competition among different ways of symmetry breaking 
generates a highly competing multicritical point. 
\par
Going back to Fig.~\ref{f1}(a), 
we see that the phase diagram has approximate reflection symmetry about the $\phi=0.5\pi$ line. 
By this reflection, the FM1 and AFM-UD1 phases are related and so as AFM-UD2 and FM2. 
The $\Gamma/J=\sqrt{3}$ tricritical point is related to another cusp point, 
and the TLL-singlet dimer transition point $J_2/J_1=0.2411, \Gamma_{\eta}=0$ ($\phi=0.075\pi,\theta=0$) 
has as the contourpart the FM-Haldane dimer transition point $J_2/J_1=-1/4$, at $\Gamma_{\eta}=0$ ($\phi=0.922\pi,\theta=0$). 
The nearly-reflection symmetry of the phase diagram is described in the Hamiltonian 
as the conversion of all the spins on one of the two legs upside down. 
Since the spin inversion is a non-unitary transformation for $S=1/2$ it is not rigorous. 
Indeed, the reflection symmetry is not perfect, particularly when $\theta\lesssim 0.25\pi$ ($\Gamma/J\lesssim1$). 
However, considering the Ising character of the magnetic phases at large $\theta$ or $\Gamma/J$, 
that join the low energy effective Hamiltonian 
near the multicritical point 
(see Appendix~\ref{app:mean}, it is natural to find that the phase boundaries of large $\theta$ are explained very well in this context.)
\par
\begin{figure*}[tbp]
    \centering
    \includegraphics[width=17cm]{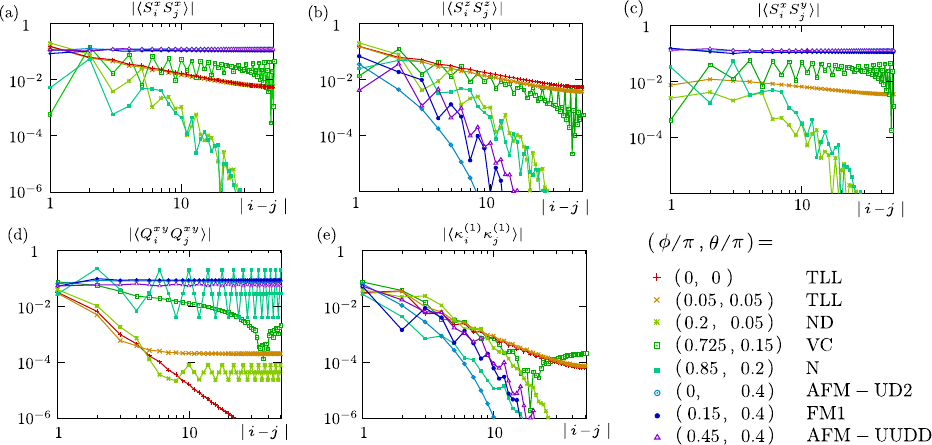}
    \caption{Two-point correlation functions $\langle O_iO_j\rangle$ 
    of (a)-(c) spins $O_i=S_i^{\alpha_i}$ of $(\alpha_i\alpha_j)=xx,zz$, and $xy$, 
    (d) nematic operators $Q_i^{xy}=S^x_i S^y_{i+1} + S^y_i S^x_{i+1}$, and 
    vector chiral operators (e) $\kappa_i=S^x_i S^y_{i+1} - S^y_i S^x_{i+1}$, 
    obtained by DMRG with $N=100$. 
    Symbols ND, VC, N, TLL indicate the nematic dimer, vector chiral, nematic, and Tomonaga-Luttinger liquid phases, respectively. 
    Several parameters $(\phi,\theta)$ are chosen from nonmagnetic and magnetic phase 
    as well as the antiferromagnetic Heisenberg chain $\phi=\theta=0$.}
    \label{f4}
\end{figure*}

\begin{figure}[tbp]
    \centering
    \includegraphics[width=8cm]{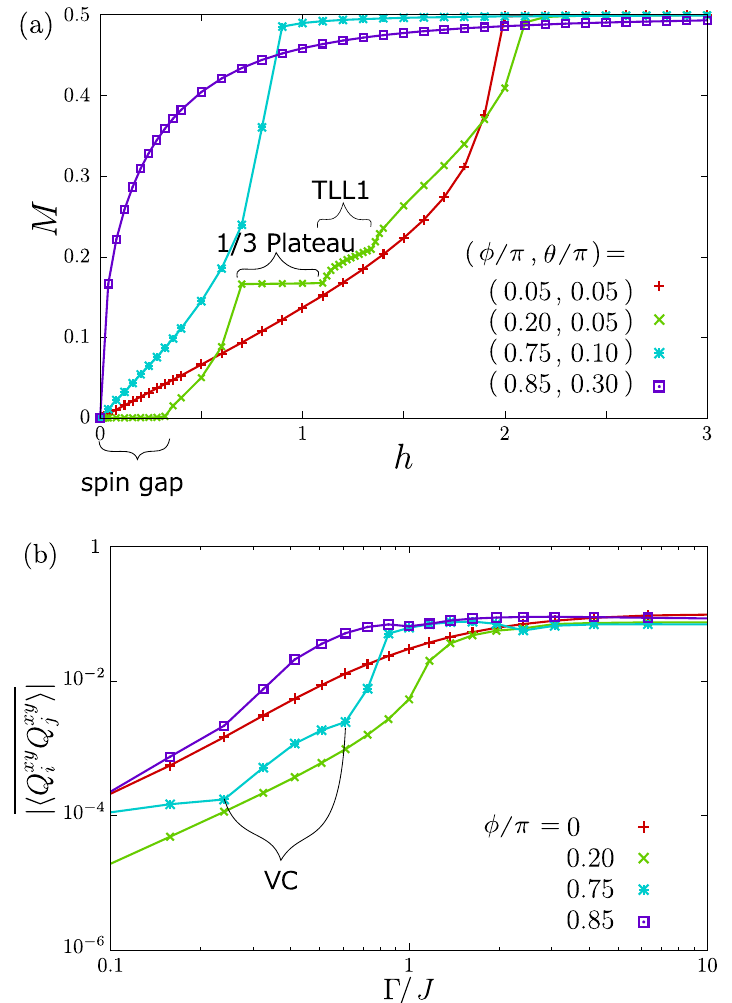}
    \caption{(a) Magnetization curves for field $h$ applied along the $z$-direction. 
     We choose $(\phi,\theta)/\pi=(0.05,0.05)$ (($J_2/J_1,\Gamma/J)\approx(0.16,0.16)$) for TLL, $(0.2, 0.05)$ ((0.72,0.16)) for ND, 
     $(0.75, 0.1)$ ((-1.0,0.32)) for VC, and $(0.85, 0.3)$ ((-0.51,1.4)) for N phases. 
     Among them, only the ND phase has a spin gap $\Delta\approx 0.3$. 
    (b) Nematic parameter $\overline{\langle Q^{xy}_i Q^{xy}_j \rangle}$ averaged over $|i-j|=58-61$ 
    as a function of $\Gamma/J$ for the choices of $\phi/\pi$ corresponding to those of panel (a). 
    All of them behave in power of $\Gamma/J$.
    }
    \label{f5}
\end{figure}
\subsubsection{Correlation functions}
We obtain the two-point correlation functions $\langle O_iO_j\rangle$,  
using the single-site or two-site operator $O_i$. 
For $O_i$, we take spin operators, $S_i^\alpha$ with $\alpha=x,y,z$, 
vector chiral operator $\kappa_i^{(n)}$, and nematic operator $Q^{xy}_i$. 
Figure~\ref{f4} shows these correlation functions as functions of distance, $|i-j|$, obtained using SSD-DMRG. 
Let us summarize the features of each phase. 
\\
{\it TLL phase.} 
Basically, the correlation functions all decay algebraically up to $|i-j|\le 50$, where we reach half of the system size. 
However, at $\theta\ne 0$, 
$\langle Q_i^{xy}Q_j^{xy}\rangle$ shows a robust constant value at large distances, unlike the standard TLL phase. 
This is because if the system remains paramagnetic without any other orderings, 
the $\Gamma$ term works as a conjugate field to induce a finite value of $\langle Q_i^{xy}\rangle$, 
which is obvious from Eq.(\ref{eq:ham}). 
In that context, the emergent $Q_i^{xy}$ is trivial as it does not break any symmetry of the Hamiltonian. 
\par
{\it Nematic-singlet dimer (ND) phase.}  
Along the exactly solvable MG line $J_2/J_1=\Gamma_2/\Gamma_1=1/2$ at $\theta < {\rm arctan}(\sqrt{3})$, 
we find the dimer-product singlet states as an exact ground state, 
which strictly excludes other components and gives $\langle Q_i^{xy}Q_j^{xy}\rangle \rightarrow 0$. 
While, even when we are away from the MG line, the ND phase sustains where 
the magnetic correlation functions all decay exponentially, indicating the existence of a spin gap. 
Such gap opening is due to the breaking of translational symmetry in the same manner as the MG solution. 
The nematic correlation $\langle Q_i^{xy}Q_j^{xy}\rangle$ starts to converge to a small 
but finite constant value at long enough distances, 
indicating the formation of a long-range order with $\Gamma_\eta\neq 0$, as shown in Fig.~\ref{f4}, 
whose implication will be discussed shortly. 
\par
{\it Vector chiral (VC) phase.}  In the FM-AFM Heisenberg interaction range of the phase diagram 
at $\phi/\pi\sim 0.7-0.8$ ($J_2/J_1= -1.3 \sim -0.72$), we find a VC phase where 
the magnetic order is absent and $\langle \kappa^{(1)}_i\kappa^{(1)}_j\rangle$ sustains at long distances. 
Its amplitude shows a large oscillation in the period of chirality over about $30$ spins as can be seen from 
the behavior of $\langle S_i^{x}S_j^{x}\rangle$ and $\langle S_i^{x}S_j^{y}\rangle$, 
while the pure magnetic components decay as a power law. 
\par
{\it Magnetically ordered phases.} 
The FM phases, AFM-UD, and UUDD phases all have their magnetic easy axis pointing in the $x'$ or $y'$ directions 
perpendicular to the leg. 
Indeed, the correlation functions show a clear exponential decay about $\langle S^z_iS^z_j\rangle$, 
while those of the in-plane elements robustly take the constant value of order-1 throughout $|i-j|$. 
The nematic correlation is also robust because these magnetic orderings are the condensation of the off-diagonal 
$S=1$ elements. 

\subsubsection{Spin gap}
Our model does not conserve total-$S^z$, in which case the spin gap cannot be evaluated by the standard treatment 
of measuring the lowest-energy difference between different total-$S^z$ sectors. 
Instead, we apply a grand canonical appoach using SSD\cite{Hotta2012,Hotta2013} that allows the access to 
the bulk magnetization curve for $N\gtrsim 20$ calculations (while we adopt $N=100$). 
By adding a Zeeman field $-h \sum_j S^z_j$ to Eq.(\ref{eq:ham}), and by deforming them with the sine-square function, 
the magnetization $M$ is obtained by extracting the intrinsic values near the center 
using the $q=0$ element of the SSD Fourier transformation\cite{Kawano2022}. 
Figure~\ref{f5}(a) shows four different magnetization curves obtained 
for the nonmagnetic ground states, TLL, ND, VC, and N. 
Only the ND phase shows a substantially large spin gap, $\Delta \sim 0.3$. 
The spin gap of $J_1-J_2$ chain ($\Gamma=0$) was previously calculated by DMRG\cite{White1996} 
and was evaluated as $\Delta=0.11\sim0.37$ for $0.14\pi<\phi<0.25\pi$ ($0.47<J_2/J_1<1.0$). 
For TLL and VC, the standard magnons condense and form a standard magnetization curve. 
Whereas, the magnetization curve of the N-phase exhibits a steep power-law increase starting from zero field, 
which reminds us of the magnetization curve of the ferrimagnetic-like state\cite{Yamashita2021}. 
Such a high sensitivity of magnetization shall appear 
because the $S=1$ and $S^z=1$ gapless excitation from the N phase may exist. 
\par
To confirm the presence of a finite spin gap, we examined the field-dependent magnetization 
in the direction of $-h \sum_j S^\alpha_j$, where $\alpha=x,y$, and $x',y'$ ($\pi/4$-rotation of $x,y$ about the $z$-axis) 
as shown in Appendix~\ref{app:spingap}; 
in the ND phase, the spin gap is finite and does not depend much on the direction of a field. 
\subsubsection{Nematic order parameter}
We have shown in Fig.~\ref{f4} that the nematic correlation, $\langle Q^{xy}_i Q^{xy}_j \rangle$, 
starts to converge to a robust constant value once we introduce finite $\Gamma/J\ne 0$. 
To examine how they behave at around $\Gamma/J\sim 0$, we plot in Fig.~\ref{f5}(b) 
the averages of $\overline{\langle Q^{xy}_i Q^{xy}_j \rangle}$ over $|i-j|=58-61$ 
to exclude the strong oscillation effect\cite{Hotta2012}. 
Numerically, there is no ``gap" in the onset value of $\big(\overline{\langle Q^{xy}_i Q^{xy}_j \rangle}\big)^{1/2}$, 
i.e. it increases immediately from $\Gamma=0$ in power of $\Gamma$ for all displayed parameters of $\phi$. 
As mentioned, $\Gamma$ in Eq.(\ref{eq:ham}) 
works as chemical potential of nonmagnetic quasi-particle 
represented by the operator $Q^{xy}_i$. 
\par
The AFM-UUDD has the staggard magnetic moment pointing in the $y'$ direction which is nothing but the 
``magnetization" represented by $Q^{xy}$. 
Namely, the doped quasi-particles condense and form a regular four-fold periodic structure breaking the 
translational symmetry. 
In the ND phase, the two-fold periodic breaking of translational symmetry occurs 
but there is no magnetic order because of the interplay of singlet and nematic particles. 
The spin gap is open, which is distinct from the AFM-UUDD and TLL phases. 
\par
For these reasons, we understand that ``field" effect, $\Gamma$, generates a finite nematic correlation 
but it appears in different ways depending on the degree of frustration dictated by $\phi$. 
As we see shortly in \S.\ref{app:dyn}, the ND phase hosts gapless excitation mode due to $\Gamma$ 
that gives the finite but small distribution of $\langle Q^{xy}_i\rangle$ in real space on top of the singlets, 
and at large $\Gamma$ the nematic particles finally condense into the magnetically ordered phases, AFM-UD2, AFM-UUDD. 
The TLL phase, although hosting finite nematic correlation, shall be trivial.

\begin{figure*}[tbp]
    \centering
    \includegraphics[width=18cm]{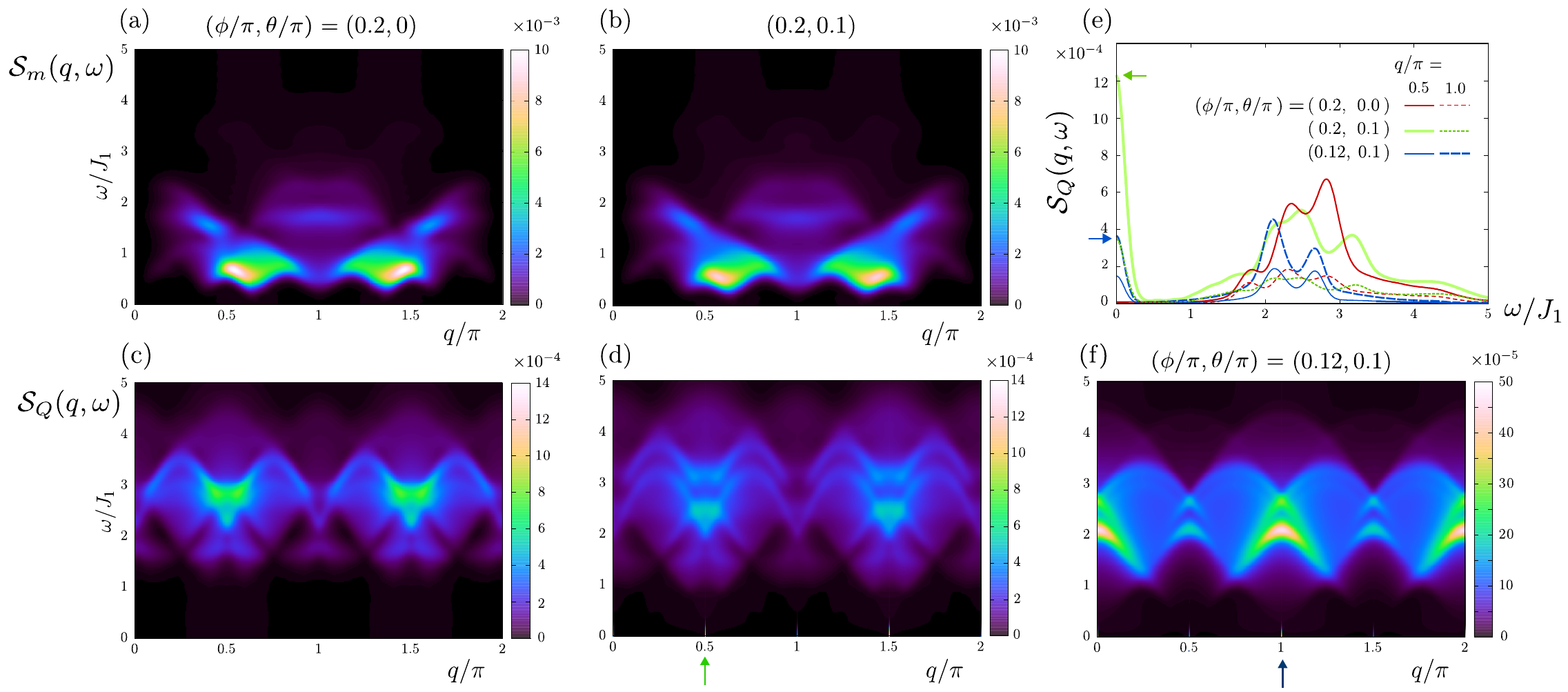} 
    \caption{(a,b) Spin and (c-f) nematic dynamical structure factors, ${\cal S}_{m}(q,\omega)$ and ${\cal S}_{Q}(q,\omega)$. 
For the density plots given in (a-d), we take $\phi/\pi=0.2$ corresponding to $J_2/J_1\approx 0.73$ with  
$\theta/\pi=0$ ($\Gamma=0$) and $0.1$($\Gamma/J\approx 0.32$) for the pure singlet dimer and ND phases, respectively. 
(e) The frequency dependences of ${\cal S}_{Q}$ are shown for fixed wave numbers, 
$q=\pi/2$ (solid line) and $\pi$ (broken line), 
where we take $(\phi/\pi,\theta/\pi)=(0.2,0),(0.2,0.1),(0.12,0.1)$. 
The green and blue arrows indicate the $q=\pi/2$ and $q=\pi$ peaks at $\omega\sim 0$, respectively.  
(f) The density plot of ${\cal S}_{Q}$ at $(\phi/\pi,\theta/\pi)=(0.12,0.1)$ ($J_2/J_1\approx0.40,\Gamma/J\approx0.32$) 
is shown which is to be compared with panel (d). 
}
    \label{fap3}
\end{figure*}
\subsection{Dynamical structure factor}
\label{app:dyn}
As we found in Fig.~\ref{f5}(b), the long-distant value of $\langle Q^{xy}_iQ^{xy}_j \rangle$ increases in power of $\Gamma/J$, 
suggesting that the nematic order parameter $\langle Q^{xy}\rangle\sim \sqrt{\langle Q^{xy}_iQ^{xy}_j \rangle}$ 
becomes finite 
by the introduction of infinitesimally small $\Gamma/J$, even for the ND phase where the singlet long-range order is present. 
When regarding $\Gamma$ as a ``field" coupled to $Q^{xy}$, this indicates that the ground state is gapless. 
To confirm it, we calculate the dynamical structure factors using the standard time-evolving block decimation (TEBD) 
technique\cite{Vidal2004} as, 
\begin{align}
{\cal S}_{m/Q}(q,\omega) &= \frac{1}{N}\sum_{j=1}^N \int \frac{dt}{2\pi}e^{i(\omega t-q r_{j})-\eta^2 t^2} G_{m/Q}(r_j,t),
\label{eq:dyn}
\\
& G_{m}(r_j,t)= 
\; \langle 0| \bm{S}_j(t)\cdot \bm{S}_{N/2}(0)|0\rangle \;, 
\notag \\
& G_{Q}(r_j,t)= 
\; \langle 0| Q^{xy}_j(t)\cdot Q^{xy}_{N/2}(0)|0\rangle ,
\label{eq:G}
\end{align} 
where $|0\rangle$ the ground state of the Hamiltonian and 
$r_{j}=j-N/2$ is the one-dimensional coordinate of site $j$ measured from the center site $N/2$. 
We consider two types of dynamical structure factors, ${\cal S}_m(q,\omega)$ and ${\cal S}_Q(q,\omega)$,  
that takes account of the magnon and nematic excitations\cite{Ramos2018, Penc2012}, respectively. 
The nematic excitations are indeed detected in resonant inelastic x-ray scattering (RIXS)\cite{Haverkort2010, Ament2011}. 
Notice that since we take the lattice spacings between $j$ and $j+1$ as a unit in the Fourier transformation 
the reciprocal number $q$ is defined accordingly. 
\par
We prepare an initial state $\bm S_{N/2}|0\rangle$ and $Q^{xy}_{N/2}|0\rangle$ and perform the 
TEBD with a timestep of $\delta t=0.1$ up to the maximum time, $T_{{\rm max}}=34\sim 37$ 
for $N=400$ with open boundary using the maximum bond dimension, $\chi=400$, 
and the Gaussian broadening $\eta^2=0.004$. 
\par
Figures~\ref{fap3}(a)-(d) show ${\cal S}_m$ and ${\cal S}_Q$ 
in the pure dimer $(\phi/\pi,\theta/\pi)=(0.2,0)$ and nematic-singlet dimer $(0.2,0.1)$ phases. 
The spin dynamical structure factors in panels (a,b) clearly show a finite gap $\Delta/J_1\sim0.2-0.3$ 
and the peak wave number $q\sim 0.7,1.3$, which is consistent with the spin gap in Figure~\ref{f5}(a) 
and the incommensurate period $q_{\text{peak}}$ in Figure~\ref{f3}(d), 
demonstrating that the single magnon dispersion is insensitive to $\Gamma$ terms. 
A strong peak structure is observed near $\omega=\Delta$ at $q=q_{\text{peak}}$, which is consistent 
with the results for the Heisenberg zigzag ladder in Ref.[\onlinecite{Sacramento2002}] at $J_2/J_1=1, \Gamma/J=0$.
\par
Regarding the nematic dynamical structure factor in Figs.~\ref{fap3}(c-f), there is a 
the distinct difference between those of zero and finite $\Gamma$; 
When $\Gamma=0$ in panel (c), the spectrum has a gap up to $\omega\sim J_1$, whereas 
introducing $\Gamma\ne 0.1$ in panel (d), 
there appears a new weight at $\omega\sim 0$ and $q=\pi/2$ indicated by an arrow. 
In Fig.~\ref{fap3}(e), the intensity of ${\cal S}_Q(\omega,q)$ for fixed $q$ is given as a function of $\omega$, 
where the $\omega=0$ peak is found to be robust.  
We confirmed that the peak positions are located at $q=\pi$ for $J_2/J_1<0.5$ and at $q=\pi/2$ for $J_2/J_1>0.5$, 
which are consistent with the ordering periods of AFM-UD2 and UUDD phases, respectively, 
that appear in the larger $\Gamma$-part of the phase diagram. 
\par
We thus find that there is a substantial gapless low energy component carried by $\langle Q^{xy}\rangle$, 
which is less dispersive and conforms to the translational symmetry broken structure of single dimers. 
This gapless weight is the precursor for the magnetic long-range orderings that take place at large $\Gamma/J$. 
\section{Bond-operator mean-field theory}
\label{sec:bond}
In this section, we examine the effect of $\Gamma$-term in the AFM-AFM zigzag chain ($J_1,J_2>0, 0\le \phi \le \pi/2$) 
by using bond-operator mean-field theory\cite{Sachdev1990}. 
We see that in the Hamiltonian Eq.(\ref{eq:ham}) 
$\Gamma$ works as a ``field" (or a chemical potential) to condense the nematic particle represented by $Q_i^{xy}$. 
This situation is similar to doping a magnon to the gapped singlet state\cite{Sachdev1990}. 
The condensation of nematic particles is discussed previously, 
where a ring exchange interaction or frustrated interactions work as a chemical potential\cite{Ueda2007,Totsuka2012,Yokoyama2018}. 
All these cases take a gapped singlet ground state as a starting point, 
which is approximated by the product state of dimers (or tetramers), 
and the effect of the applied ``field" is examined using the bond-operator approach or its analog. 
\par
\begin{figure*}
    \centering
    \includegraphics[width=17cm]{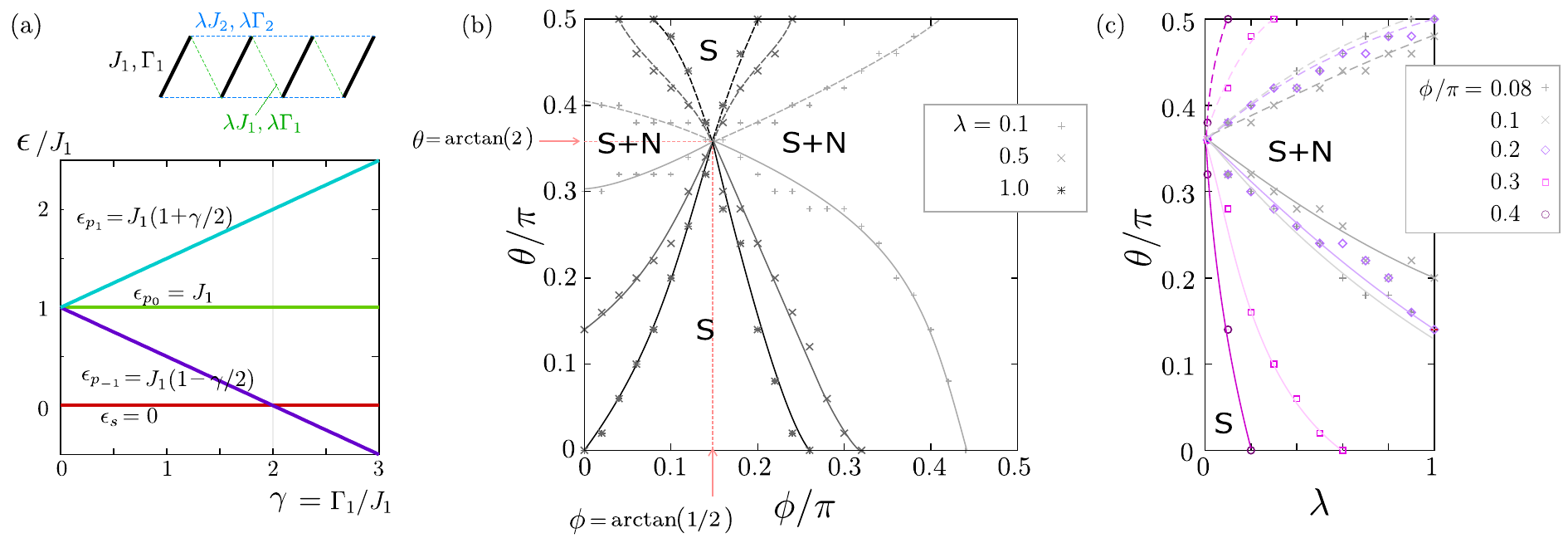}
    \caption{(a) Zigzag lattice based on the dimer unit, where the inter-dimer interactions scaled by $\lambda$, 
             and the eigenenergy levels of a single isolated dimer 
             as a function of $\Gamma_1/J_1 \equiv \gamma$. 
     Phase diagrams obtained by the bond-operator mean-field theory 
     (b) $\theta-\phi$ plane for $\lambda=0.1,0.5,1$, and (c) $\theta-\lambda$ plane for several choices of $\phi$. 
     Symbols S, S+N indicate the singlet$(\bar{s}^2=1)$ and the nematic-singlet phase$(\bar{s}^2<1)$.
}
    \label{f6}
\end{figure*}

We consider a model shown in Fig.~\ref{f6}(a) consisting of dimers (rungs) 
where the intra-dimer interaction is $J_1,\Gamma_1$ and the inter-dimer interactions 
are $\lambda J_\eta,\lambda \Gamma_\eta,\; (\eta=1,2)$ with $0\leq \lambda \leq 1$. 
The $\lambda=1$ limit corresponds to Eq.(\ref{eq:ham}). 
We start from $\lambda=0$ having decoupled dimers, whose ground state is the product state 
of the lowest energy isolated-dimer states. 
By using the three excited eigenstates of the dimer represented by the three bond operators, 
we rewrite the Hamiltonian including the nonzero $\lambda$-terms 
and examine the nature of the low-energy excitations 
by the bond-operator mean-field theory. 
%

\subsection{Single dimer}
\label{sec:bond1}
The Hamiltonian of a single dimer is $\hat h=J_1(\bm {S}_1\cdot \bm {S}_2+\gamma(S_1^xS_2^y+S_1^yS_2^x))$, 
where we parameterize $\gamma=\Gamma_1/J_1=\Gamma_2/J_2$. 
Its eigenstates are given as 
\begin{align}
    &|s\rangle = (|\uparrow \downarrow \rangle - |\downarrow \uparrow \rangle)/\sqrt{2}, \notag\\
    &|p_0\rangle = (|\uparrow \downarrow \rangle + |\downarrow \uparrow \rangle )/\sqrt{2},\notag\\
    &|p_1\rangle = (|\uparrow \uparrow \rangle + i|\downarrow \downarrow \rangle )/\sqrt{2},\notag\\
    &|p_{-1}\rangle = (|\uparrow \uparrow \rangle - i|\downarrow \downarrow \rangle )/\sqrt{2},
    \label{eq:H_rung_eigenvectors}
\end{align}
and the corresponding eigenvalues as $\epsilon_s= 0, \epsilon_{p_0}=J_1$, and 
$\epsilon_{p_{1/-1}}=J_1(1\pm\gamma/2)$ 
(see Fig.~\ref{f6}(a)), where we dropped the constant term $-3J_1/4$ for simplicity. 
At $\gamma=0$, the excited states are three-fold degenerate.  
The spin-singlet ground state $|s\rangle$ shows a level crossing at $\gamma=2$ and 
$|p_{-1}\rangle$ replaces the ground state. 
\par
Now, we introduce the boson operators on an $n$-th dimer, $p_{\alpha,n}$, $\alpha=x,y,z$, together with the 
singlet operator $s_n$, which describes the four eigenstates as 
\begin{align}
   & |s\rangle_n \;\;= s_n^\dagger |0\rangle ,\notag\\
   & |p_0\rangle_n = \;p_{z,n}^{\dagger}|0\rangle ,\notag\\
   & |p_1\rangle_n = \;\frac{e^{i3\pi/4}}{\sqrt{2}}(p_{x,n}^{\dagger}-p_{y,n}^{\dagger})|0\rangle,\notag\\
   & |p_{-1}\rangle_n = \frac{e^{-3\pi/4}}{\sqrt{2}}(p_{x,n}^{\dagger}+p_{y,n}^{\dagger})|0\rangle,
    \label{eq:p_-101}
\end{align}
where $|0\rangle$ is a vacuum. 
These operators satisfy the local constraint 
\begin{align}
    s_n^{\dagger}s_n+\sum_{\alpha}p_{\alpha,n}^{\dagger}p_{\alpha,n}=1. 
    \label{eq:par_num}
\end{align}
and the commutation relations, 
$[s_n,s_{n'}^{\dagger}] = \delta_{nn'}, \quad 
[p_{\alpha,n},p_{\beta,n'}^{\dagger}]=\delta_{\alpha \beta}\delta_{nn'},\quad 
[s_n,p_{\alpha,n'}^{\dagger}]=0$, {\rm etc.}

\subsection{Effect of inter-dimer interaction}
\label{sec:bond2}
We start from $\lambda=0$ and $0<\gamma<2$, whose ground state is given as $\otimes_{n=1}^{N/2}|s\rangle_n$. 
We consider the system consisting of $N_{\mathrm{d}}=N/2$ dimers with periodic boundary. 
The spin operators are described in terms of boson operators on the $n$-th dimer ($p_{\alpha,n}$) as 
\begin{align}
S_1^{\alpha} &= \frac{1}{2}(s_n^{\dagger}p_{\alpha,n} + p_{\alpha,n}^{\dagger} s_n - i \epsilon_{\alpha \beta \gamma} p_{\beta,n}^{\dagger} p_{\gamma,n}),
    \notag \\
S_2^{\alpha} &= \frac{1}{2}(-s_n^{\dagger}p_{\alpha,n} - p_{\alpha,n}^{\dagger} s_n - i \epsilon_{\alpha \beta \gamma} p_{\beta,n}^{\dagger} p_{\gamma,n}),
    \label{eq:bos_rep}
\end{align}
and using them, the interaction terms between the $n$ th and $(n+1)$ th dimer, 
consisting of three bonds are rewritten in the form of two and four body terms, 
as shown in Appendix \ref{app:bop}, Eq.(\ref{eq:S_com}). 
\par
Here, we consider a small $\lambda$, assuming $\gamma<2$ ($\theta<0.35\pi$), 
expecting that the ground state is dominated by singlets and the population of $p_{\alpha,n}$-bosons are dilute. 
The condensed singlets are replaced by its expectation value $\langle s \rangle = \bar{s}$, 
and by dropping off the terms consisting of three or four $p$ operators, 
we obtain the interactions between the $n$-th and $(n+1)$-th dimers as  
\begin{align}
    \hat h_n^{\rm MF} 
    &= \frac{-J_1+2 J_2}{4} \bar{s}^2 \big\{(p_{\alpha,n}+p_{\alpha,n}^{\dagger})
       (p_{\alpha,n+1}+p_{\alpha,n+1}^{\dagger}) \notag \\ 
    & \quad + \gamma (p_{x,n}+p_{x,n}^{\dagger}) (p_{y,n+1}+p_{y,n+1}^{\dagger}) \notag \\
    & \quad + \gamma (p_{y,n}+p_{y,n}^{\dagger}) (p_{x,n+1}+p_{x,n+1}^{\dagger}) \big\}. 
    \label{eq:interrung}
\end{align}

\subsection{Bond-operator mean field Hamiltonian}
\label{sec:hbop}
To fulfill the condition Eq.(\ref{eq:par_num}) we introduce the Lagrange multiplier $\mu$ common to all dimers, 
assuming the translation invariance of the system in a unit of dimer, 
and add the chemical potential ($\mu$) term to the Hamiltonian. 
Then, we finally reach the form of the mean-field Hamiltonian 
consisting only of bilinear terms as 
\begin{align}
    {\cal H}_{\mathrm{bo}} &= \sum_{n=1}^{N_{\mathrm{d}}} ( \hat h_n^{\rm d} +  \hat h_n^{\rm MF} ) \notag\\
    & \hat h_n^{\rm d}= J_1 \Big(\sum_{\alpha=x,y,z} p_{\alpha,n}^{\dagger}p_{\alpha,n} + 
       \gamma/2 ( p_{x,n}^{\dagger}p_{y,n} + p_{y,n}^{\dagger}p_{x,n} )  \Big)\notag \\
    &\quad\quad  -\mu \big(\bar{s}^2 + \sum_{\alpha=x,y,z}p_{\alpha,n}^{\dagger}p_{\alpha,n} -1 \big). 
    \label{eq:H_bo}
\end{align}
Here, the mean-field parameters $(\mu,\bar{s}) \in {\mathbb R}$ are determined within the physically meaningful range 
to minimize $\langle {\cal H}_{\mathrm{bo}} \rangle$. 
\par
By performing a Fourier transformation, 
$p_{\alpha,n}^{\dagger} = N_{\mathrm{d}}^{-1/2}\sum_k e^{ik n}p_{\alpha,k}^{\dagger}$, 
we obtain the form, 
\begin{align}
   {\cal H}_{\mathrm{bo}} = \sum_k \{\bm{u}_k^{\dagger} A(k) \bm{u}_k + \bm{v}_k^{\dagger} B(k) \bm{v}_k\}
    \notag \\
    +N_{\mathrm{d}}\Big(\frac{5}{2}\mu - \frac{3}{2}J_1 - \mu \bar{s}^2\Big),
    \label{eq:H_bo_compact}
\end{align}
where $\bm{u}_k = (p_k^x,p_k^y,p_{-k}^{x\dagger},p_{-k}^{y\dagger})^T,\ \bm{v}_k = (p_k^z,p_{-k}^{z\dagger})^T$, 
and $A(k)$, $B(k)$ are the $4\times 4$ and $2\times 2$ real symmetric matrices (see Appendix~\ref{app:bop}).
\par
By performing a Bogoliubov transformation, we can diagonalize ${\cal H}_{\mathrm{bo}}$, which then yields 
\begin{align}
   {\cal H}_{\mathrm{bo}} &= \sum_k \sum_{l=1}^3 \omega_l (k) \tilde p_{k,l}^{\dagger} \tilde p_{k,l} 
     +E_{\mathrm{GS}}^{\mathrm{MF}},
    \label{eq:H_bo_dis}
\end{align}
where the dispersion relations and the mean-field ground-state energy are given as
\begin{align} 
    \omega_1(k) &= \Big[\big(\epsilon_{p_{1}}-\mu\big)\big(\epsilon_{p_{1}}-\mu +\lambda(1-\gamma) \varepsilon(k) \big)\Big]^{1/2},
    \notag\\
    \omega_2(k) &=\Big[(\epsilon_{p_{-1}}-\mu)\big( (\epsilon_{p_{-1}}-\mu) +\lambda(1+\gamma)\varepsilon(k) \big)\Big]^{1/2},
    \notag\\
    \omega_3(k) &=\Big[(\epsilon_{p_0}-\mu )\big(\epsilon_{p_0}-\mu+\lambda \varepsilon(k) \big)\Big]^{1/2}, 
\notag\\
   & \quad\varepsilon(k)= (-J_1+2J_2)\bar{s}^2\cos k. 
    \label{eq:wdispersion}
\end{align}

\par
\begin{figure}
    \centering
    \includegraphics[width=8cm]{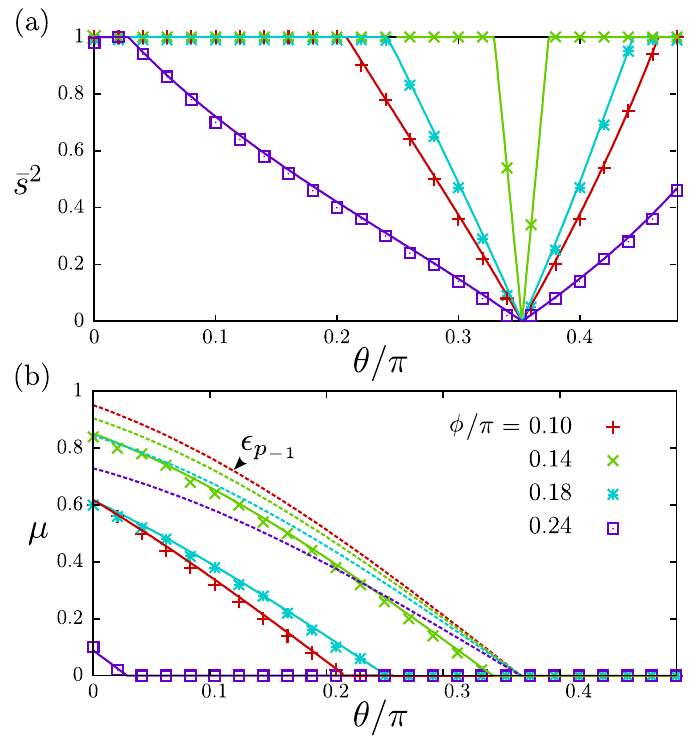}
    \caption{The rate of singlet particle condensation $\bar{s}^2$ and the chemical potential $\mu$ with $\lambda=1$.}
    \label{f7}
\end{figure}
\par
\begin{figure}
    \centering
    \includegraphics[width=9cm]{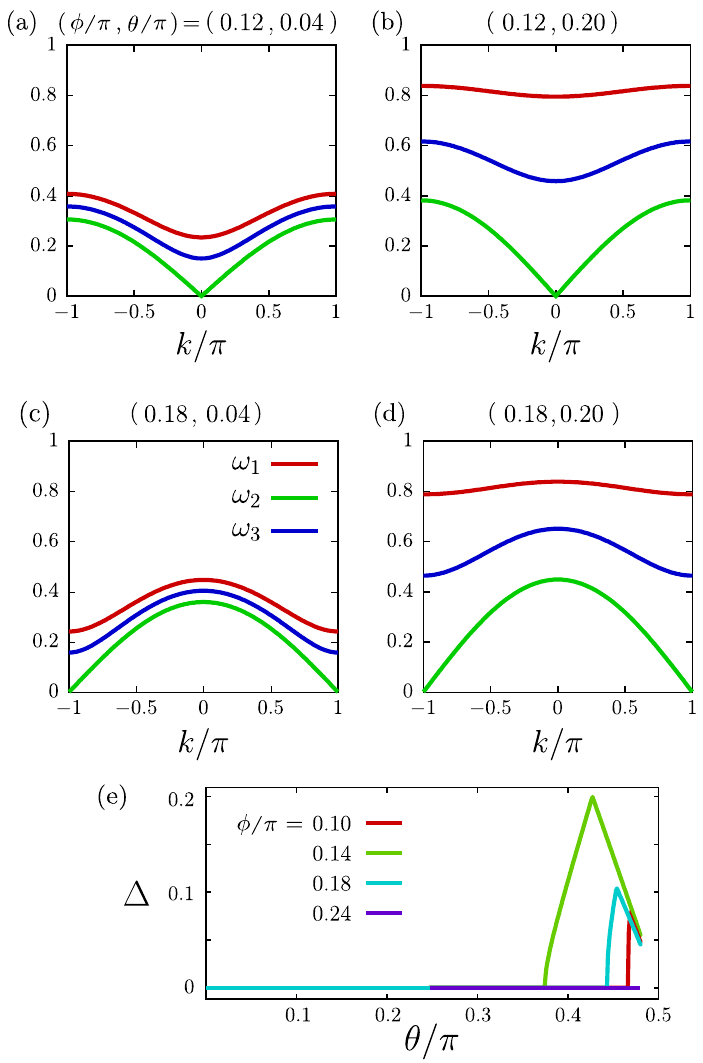}
    \caption{(a)-(d) Dispersion relation, $\omega_l$, $l=1,2,3$ of 
    ${\cal H}_{\rm bo}$ obtained for several choices of $\phi,\theta$. 
    (e) Energy gap $\Delta$ of ${\cal H}_{\rm bo}$ normalized by $\sqrt{J_1^2+J_2^2+\Gamma_1^2+\Gamma_2^2}$ as functions of $\theta$ 
    for four choices of $\phi$. }
    \label{f8}
\end{figure}

\subsection{Phase diagram}
\label{sec:bond3}
The phase diagrams obtained by the bond-operator mean-field theory are shown in 
Figs.~\ref{f6}(b) and \ref{f6}(c) for several fixed values of 
$\lambda=0.1,0.5,1.0$ and $\phi/\pi=0.08$ to $0.4$ ($J_2/J_1=0.26-3.1$), respectively. 
The corresponding set of mean-field parameters $(\mu,\bar s^2)$ are shown in Fig.~\ref{f7} at $\lambda=1$ 
as functions of $\theta$. 
For these parameters, there is a range of small $\theta$ that the ground state is fully occupied by singlets, 
$\bar s^2=1$, which we call S-phase. 
The phase transition from S-phase to (S+N)-phase takes place at finite $\theta$, 
where $\bar s^2$ starts to decrease from 1, and the chemical potential $\mu$ reaches the singlet (zero) energy level. 
When increasing $\lambda$, the S-phase shrinks because 
the $p$-particles gain the kinetic energy due to inter-dimer interactions, 
but at $\phi={\rm atan}(1/2)$, i.e. $J_2/J_1=\Gamma_2/\Gamma_1=0.5$, 
which corresponds to the Majumdar-Ghosh line, the singlet phase remains stable throughout 
the whole range of $\theta$. 
This fact is trivial as we find that the inter-dimer interaction in 
Eq.(\ref{eq:interrung}) is $\hat h_n^{\rm MF}=0$, 
or equivalently, the dispersion in Eq.(\ref{eq:wdispersion}) yields $\epsilon(k)=0$ for $J_1=2J_2$. 
Indeed, along this line, the singlet product state is the exact solution of 
the original Hamiltonian Eq.(\ref{eq:ham}) without any approximation. 
\par
We plot in Figs.~\ref{f8}(a)-\ref{f8}(d) the energy dispersions $\omega_l(k)$ at $\lambda=1$ for several choices of 
$\phi$ and $\theta$, inside the S-phase. 
Remarkably, all of them are gapless. 
Indeed, the energy gap $\Delta$ of ${\cal H}_{\rm bo}$ shown in Fig.~\ref{f8}(e) 
is zero for all parameters of $\phi$ at small $\theta$. 
This shows that the singlet phase is robust but at the same time, the $p$-particle 
are always able to join the ground state. 
This result is rather unusual; 
the standard bond operator approach is used to provide the instability toward 
the condensation of these excited $p$-state that replaces the ground state, 
and the gap-closing point usually indicates the phase transition. 
Whereas, in the present case, the zero gap of ${\cal H}_{\rm bo}$ does not mean the instability of the singlet ground state. 
We emphasize that the instability takes place not at the gap closing point but when $\bar s^2$ starts to deviate from 1, 
and the chemical potential $\mu$ reaches the singlet energy level. 
For clarification, we plot in Fig.~\ref{f7}(b) together the bare excited energy, $\epsilon_{p_{-1}}$, 
of the isolated dimer. We see that $\mu$ is located in between $\epsilon_s$ and $\epsilon_{p_{-1}}$. 
This result is consistent with numerical results in the previous section; 
the coexistent nematic and singlet orders that break the translational symmetry sets in immediately when $\theta>0$, 
which corresponds to the S-phase. 
\par
The $\omega_2$ branch in Fig.~\ref{f8} has a gapless point 
at $k=0$ ((a,b) $J_2/J_1<0.5$) and $k=\pi$ ((c,d)$J_2/J_1>0.5$) for periods of dimer. 
They agree with the gapless nonzero weight 
of the dynamical structure factor $q=\pi$ ($J_2/J_1<0.5$) and $q=\pi/2$ ($J_2/J_1>0.5$) 
defined in the unit of lattice sites in Fig.~\ref{fap3}, respectively. 
When $\theta$ reaches the phase boundary, 
the whole $\omega_2$ branch condenses and replaces the singlet, 
and the magnetic long-range order in a period of the dimer (two sites) and two dimers (four sites) 
appear for $J_2/J_1<0.5$ and $>0.5$, respectively. 
Indeed, the phase diagram obtained by DMRG in Fig.~\ref{f1}(a) has AFM-UD2 and AFM-UUDD phases, respectively, 
which are the states corresponding to S+N of different periodicity. 
\par
We may interpret that the dispersiveness of $\omega_2$ around the gapless point shall be the artifact of the 
mean-field approximation to one-body 
because the higher order interactions and many-body effects are discarded (see Appendix \ref{app:bop}). 
Due to the correlation effects the weight of dispersions off the gapless point shall be weakened to form 
the gapless peak in the dynamical structure factor in Fig.~\ref{fap3}.  
\par
The $p_{1,n}^\dagger$ and $p_{-1,n}^\dagger$ are the creation operator on the $n$-th dimer 
in Eq.(\ref{eq:par_num}) of the states that give $\langle Q^{xy}_n\rangle=1$ and $-1$, respectively. 
Therefore, we evaluated the number operators 
with respect to the eigenstate $|k,2\rangle$ of $\omega_2(k)$, 
finding that 
\begin{align}
&\langle k,2| p_{1,n}^\dagger p_{1,n} |k,2\rangle=0, 
&\langle k,2| p_{-1,n}^\dagger p_{-1,n} |k,2\rangle=2/N, 
\end{align}
for all $k$. 
This indicates that each $k$-point carries a single $p_{-1,n}$ particle. 
The gapless excitation that contributes to the ground state as fluctuations 
thus yield the dilute order-$1/N$ concentration of $p_{-1,n}$, 
in agreement with the DMRG value of $\langle Q^{xy}_i\rangle \sim 0.01$ in Figs. \ref{f4}(d) \ref{f5}(b). 
\begin{figure}
    \centering

    \includegraphics[width=7.5cm]{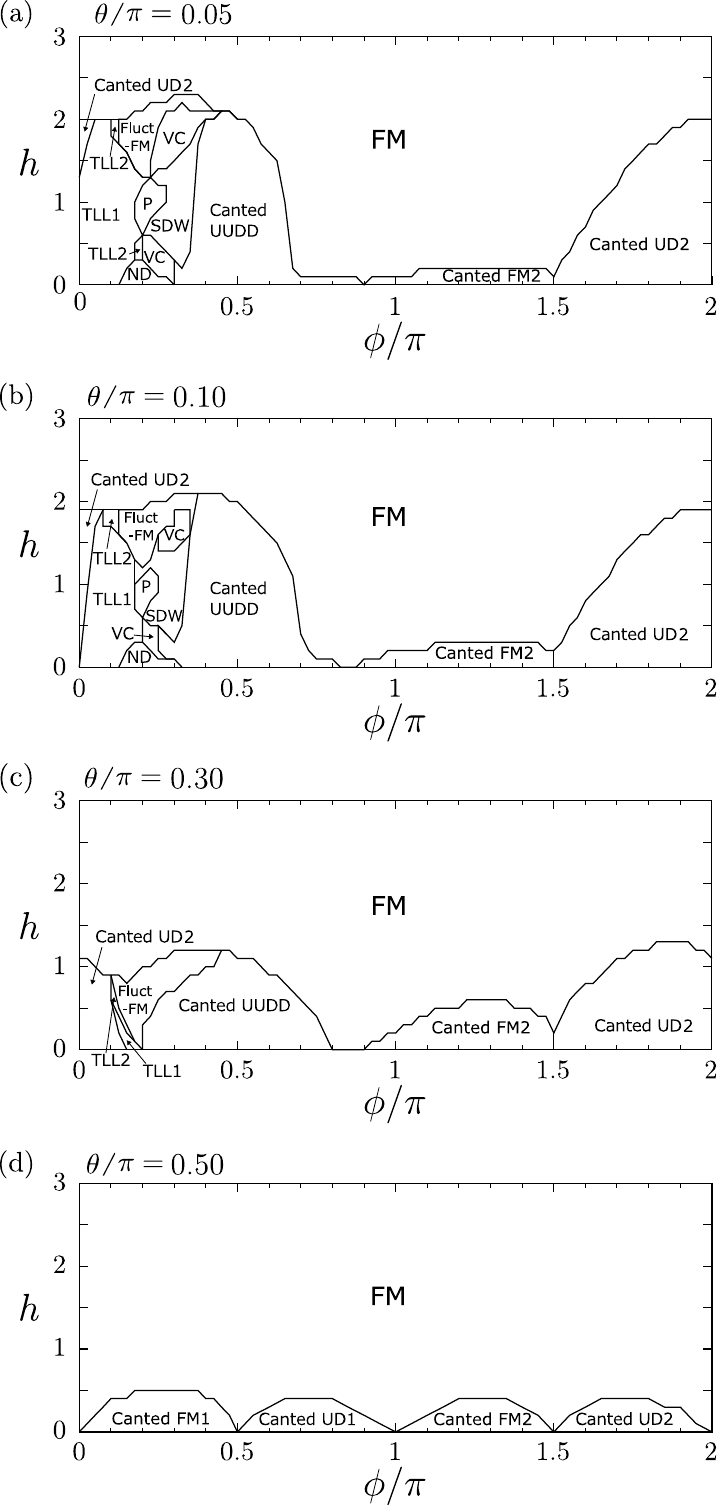}
    \caption{Ground state phase diagram in an applied field $h$ at 
            (a)-(d) $\theta=0.05\pi,0.1\pi,0.3\pi,0.5\pi$ ($\Gamma/J\approx0.16,0.32,1.4,\infty$). 
           We find ferromagnetism(FM), up down(UD), fluctuating-ferro(Fluct), vector chiral(VC), 
           Tomonaga-Luttinger liquid (TLL),spin-density wave (SDW) phases, 
           1/3-plateau(P), and nematic-singlet dimer(ND). 
}
    \label{f9}
\end{figure}
\par
\begin{figure}
    \centering
    \includegraphics[width=8.5cm]{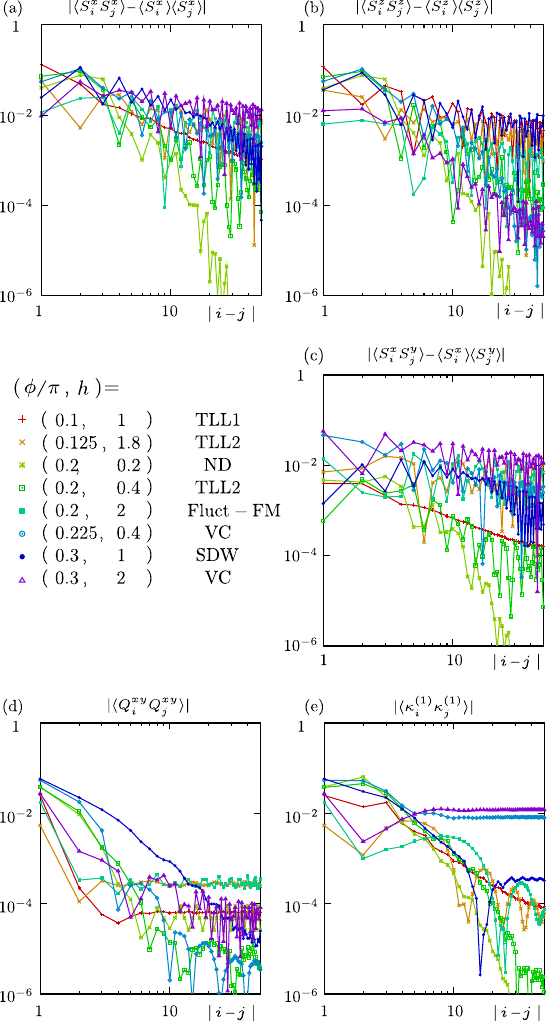}
    \caption{Two-point correlation functions $\langle O_iO_j\rangle$ 
    of nematic operators $O_i=S^x_i S^y_{i+1} + S^y_i S^x_{i+1}$, vector chiral operators $O_i=S^x_i S^y_{i+1} - S^y_i S^x_{i+1}$, 
    and $\langle O_iO_j\rangle - \langle O_i\rangle \langle O_j\rangle$ of spins 
     $O_i=S_i^{\alpha_i}$ of $(\alpha_i\alpha_j)=xx,zz$, and $xy$, 
    obtained by DMRG with $N=100$ with $\theta=0.05\pi$ ($\Gamma/J\approx0.16$). Symbols ND, VC, TLL, SDW, and Fluct-FM 
      indicate the nematic dimer, vector chiral, Tomonaga-Luttinger liquid, 
      spin-density wave and fluctuating-Ferromagnetic phases, respectively. Several parameters are chosen from nonmagnetically ordered  
      or fluctuating FM phase.}
    \label{f10}
\end{figure}
\par
\begin{figure}
    \centering
    \includegraphics[width=8cm]{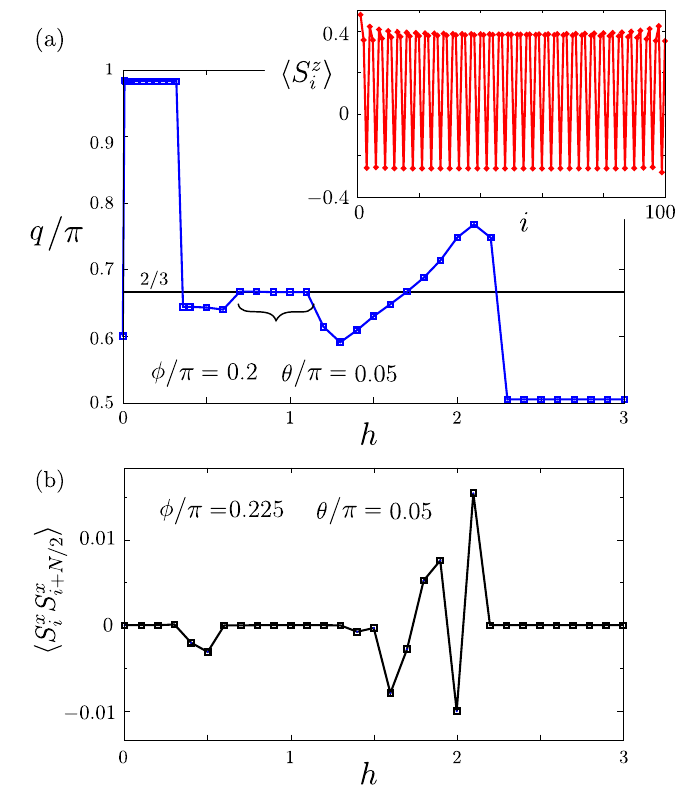}
    \caption{(a) Local magnetization $\langle S^z_i\rangle$ (upper panel) in the 1/3-plateau phase, 
    clearly showing the up-up-down spin configuration. 
    The peak wave number $q$ of ${\cal S}^{zz}(q)$ is shown as a function of $h$, 
    where, within $0.5\pi \leq k \leq \pi$ we find $q=2\pi/3$. 
    (a) The spin correlation function, $\langle S^x_{i} S^x_{i+N/2}\rangle\neq0$ ($i=25$), 
    as a function of $h$ that takes finite value in the fluctuating-FM phase at $1.5\lesssim h \lesssim 2.15$. 
}
    \label{f11}
\end{figure}
\section{Effect of magnetic field}
\label{sec:mag}
It is known that the $J_1$-$J_2$ Heisenberg model in the AFM-AFM and FM-AFM regions 
exhibits a rich phase diagram in an applied magnetic field\cite{Hikihara2008,Hikihara2010}. 
Here, our $\Gamma$-terms change these phase diagrams a lot. 
We consider an external magnetic field along the $z$-axis whose Hamiltonian is given as ${\cal H}+{\cal H}_f$ 
using Eq.(\ref{eq:ham}) with 
\begin{equation}
    {\cal H}_f=-h\sum_j S^z_j. 
     \label{eq:ham_mag}
\end{equation}
\subsection{Magnetic phase diagram} 
Figure \ref{f9} shows the magnetic phase diagram at fixed values of $\theta=0.05\pi,0.1\pi,0.3\pi,0.5\pi$ ($\Gamma/J\approx0.16,0.32,1.4,\infty$). 
We first focus on small $\Gamma$, namely $\theta=0.05\pi,0.1\pi$. 
Compared to the zigzag Heisenberg model, 
the $\Gamma$-term transforms some of the nonmagnetic phases to magnetic ordered ones. 
For example, 
the fluctuating-FM phase at around $(\phi,\theta,h)\sim (0.2,0.05,1.8)$ ($J_2/J_1\approx0.73,\Gamma/J\approx0.16$) in Fig.~\ref{f9}(a) 
was originally a TLL2 phase\cite{Hikihara2010} but was stabilized as an emergent small magnetic order in the $xy$ plane. 
Furthermore, in the FM-AFM region $0.5\pi<\phi<\pi$, 
the canted UUDD phase where up-up-down-down magnetic order emerges in the $xy$ plane, 
was originally a vector chiral, nematic, and other multipolar phases when $\Gamma=0$.  
\par
At large $\Gamma$, i.e. $\theta=0.3\pi,0.5\pi$ in Figs.~\ref{f9}(c) and \ref{f9}(d), 
the nonmagnetic phases like TLL and ND are almost wiped out. 
Magnetic phases, FM and UD, that appeared in the $h=0$ phase diagram are preserved while 
the moments are canted off the $xy$-plane because of $h$. 
\par
\subsection{Magnetic structures} 
We now summarize the features of each phase based on the two-point correlation functions 
shown in Fig.~\ref{f10}. 
\par
{\it TLL1 and TLL2 phases.}  
TLL1 phase is adiabatically connected to the TLL at $h=0$. 
In both TLL1 and TLL2, the spin-spin correlations $\langle S^\alpha_i S^{\alpha'}_j \rangle$ show algebraic decay. 
The difference from those of the pure Heisenberg case is a finite $\langle Q^{xy}_iQ^{xy}_j\rangle \ne 0$ 
at long distances because of $\Gamma\ne 0$.  
The structure factor ${\cal S}^{xx}(q)$ and ${\cal S}^{zz}(q)$ of TLL2 show peaks at incommensurate wave numbers, 
whereas that of ${\cal S}^{xx}(k)$ in TLL1 is commensurate. 
It was shown that TLL2 phase with $\Gamma=0$ is described as two Gaussian conformal field theory with central charge $c=1+1$\cite{Hikihara2010}. 
\par
{\it SDW phase.}  At large $J_2/J_1$ $(\phi\approx0.3\pi)$, 
there is a phase characterized by the incommensurate and quasi-long-ranged longitudinal correlation, 
$\langle S^z_i S^z_j\rangle - \langle S^z_i \rangle \langle S^z_j \rangle$, 
and the short-ranged transverse correlation $\langle S^x_i S^x_j\rangle$. 
Here, we find $\langle Q^{xy}_iQ^{xy}_j\rangle \rightarrow 0$, meaning that this type of nematic order does not exist. 
In the pure Heisenberg case $\Gamma=0$, 
the magnetization shows a stepwise structure by $\Delta S^z_{{\rm tot}}=2$, 
and the exponential decay of $\langle S^x_i S^x_j\rangle$ suggests a finite energy gap to single-spin-flip excitations; 
they are the signature of the nematic state. 
However, these features are no longer observed at $\Gamma >0$. 
\par
{\it 1/3-plateau phase.}  In Figs.~\ref{f9}(a) and \ref{f9}(b), we find a 1/3-plateau phase 
at around $\phi/\pi\sim 0.2$ ($J_2/J_1\sim0.73$), which has a typical up-up-down configuration of spins. 
Figure~\ref{f11}(a) shows the wave number $q$ at which the Fourier transform of $\langle S^z_j\rangle$ shows a peak. 
In our analysis, we applied the SSD Fourier transform 
$\langle S^z_q\rangle =\sum_{j=1}^N f(r_j)\langle S^z_j \rangle e^{-iq r_j}/\sum_{j=1}^N f(r_j)$ \cite{Kawano2022}, 
successfully detecting the plateau region yielding $q=2\pi/3$. 
\par
{\it VC and ND phases.}  These phases are basically the same as $h=0$. 
It is also common to the case of $\Gamma=0$ except that $\langle Q^{xy}_iQ^{xy}_j\rangle \ne 0$. 
The VC phase at $\Gamma=0$ is a one-component TLL\cite{Hikihara2010}. 
The ND phase sustains up to finite $h$ whose value corresponds to the spin gap at $h=0$. 
\par
{\it Fluctuating FM phase.} The fluctuating FM phase has a ferromagnetic order with its moments pointing in the $z$ direction 
and the small fluctuation in the $xy$ plane. 
Its origin is essentially a $\Gamma$ term because it vanishes at $\Gamma=0$. 
We determined the boundary of this phase by $\langle S^x_i S^x_{i+N/2}\rangle \ne 0$ (see Fig.~\ref{f11}(b)). 
\par
{\it Canted magnetic phases.} 
The FM1, FM2, UD1, UD2, and UUDD phases at $h=0$ developed magnetic orders in the $xy$-plane. 
In a magnetic field, these moments cant in the $z$-directions. 

\section{Discussion and Summary}
\label{sec:summary}
We have elucidated the ground state phase diagram of the zigzag Heisenberg-$\Gamma$ chain across regions of 
ferromagnetic, antiferromagnetic, and mixed couplings. 
Furthermore, we have depicted the magnetic phase diagram in an applied magnetic field oriented 
along the chain perpendicular to the $\Gamma$-type fluctuations. 
\par
In the regime where the $\Gamma$-term dominates, the model exhibits strong magnetic anisotropy, 
resulting in ferromagnetic (FM) and antiferromagnetic (AFM-UD, UUDD) orderings with moments aligned along $(x,y,z)=(1,\pm 1,0)$ directions, 
namely, in-plane and perpendicular to the chain. The application of a magnetic field tilts these moments off the plane.
\par
In contrast, for small $\Gamma$, we observe a nematic-singlet dimer (ND) phase characterized 
by long-range nematic correlations and a finite spin gap. 
This phase features an additional nonmagnetic gapless excitation arising from the nematic correlations, 
as clarified by the bond-operator mean-field theory, consistent with numerical results from DMRG simulations. 
The emergence of the ND phase stems from the intricate interplay between strong geometrical frustration and the $\Gamma$-term; 
the formation of spin singlets breaks lattice translational symmetry and impedes dominant nematic orders, 
while robust nematic correlations develop due to the nonmagnetic gapless nature. 
Ultimately, the ground state is predominantly governed by the presence of singlets. 
At larger and finite $\Gamma$, condensation of nematic $S=1$ particles leads to magnetic long-range order, 
with spins oriented in the $xy$ plane.
\par
We now explore the relevance of our findings to the ytterbium-based rare-earth magnet, YbCuS$_2$. 
The Yb ion in this material is surrounded by S$_6$ octahedra\cite{Gulay2005}, 
leading to the formation of a Kramers $\Gamma_6$ doublet effectively carrying spin-1/2 due 
to the interplay between octahedral crystal fields of S-ions and strong spin-orbit coupling interactions \cite{Ohmagari2020}. 
The crystal structure of YbCuS$_2$ is orthorhombic with space group $Pnma$ \cite{kaneshima2021}, 
where octahedra share edges along the $a$-direction and every two octahedra share edges along the $c$-direction, 
forming a one-dimensional zigzag chain of Yb ions, 
whose details are shown in Ref.[\onlinecite{Saito2024}]. 
Experimental observations of the magnetic susceptibility $\chi(T)$ reveal antiferromagnetic interactions among the spins. 
At low temperatures, YbCuS$_2$ undergoes a first-order transition at $T_O=0.95$ K, characterized by a 
pronounced divergence in the specific heat\cite{Ohmagari2020}. 
In the low-temperature phase, incommensurate magnetic structures are detected through $^{63/65}$Cu-nuclear 
magnetic resonance (NMR) and nuclear quadrupole resonance (NQR) measurements on polycrystalline samples \cite{Hori2023}. 
The nuclear spin-lattice relaxation rate $1/T_1$ of $^{63/65}$Cu-NQR exhibits a $T$-linear behavior at $T<0.5$ K, 
suggesting the presence of gapless excitations \cite{Hori2023}. 
Additionally, the magnetic-field-temperature phase diagram obtained experimentally shows a transition of 
the low-temperature phase to an up-up-down (UUD) phase at around 5 T\cite{Ohmagari2020}.
\par
The 1/3-plateau with up-up-down (UUD) magnetic structure is a common feature observed in triangular-based magnets, 
including NaYbSe$_2$ \cite{Ranjith2019,Scheie2023}, CsYbSe$_2$ \cite{Xie2023}, RbYbSe$_2$ \cite{Xing2021}, 
and KYbSe$_2$ \cite{Xing2021,Scheie2023}. 
This state is typically explained by the two-dimensional $S=1/2$ XXZ model \cite{Yamamoto2014}.
However, in the spin-1/2 zigzag Heisenberg chain, 
the parameter values required for the appearance of the robust 1/3-plateau 
in a finite field at $h\sim J_1$ is $J_2/J_1\sim 0.5-1$ that lead to a ground state dominated by the 
dimer singlet state in the absence of a magnetic field. 
This dimer singlet is fully gapped both magnetically and nonmagnetically\cite{Hikihara2010}. 
Consequently, this model alone cannot explain the observed nonmagnetic gapless behavior in the material.
\par
The motivation to understand the nonmagnetic gapless behavior in YbCuS$_2$ led us to derive 
the quantum spin model of this material using perturbation theory \cite{Saito2024}. 
Specifically, we evaluated the super-exchange interactions between Yb spins mediated by S-ions through fourth-order processes. 
Our analysis revealed that the diagonal exchange interaction is nearly Heisenberg-like, 
with a small off-diagonal $\Gamma$-type term. 

The presence of this off-diagonal anisotropic interaction can be attributed to two main factors. 
Firstly, the large splitting of $f^{12}$ states selects a particular orbital momentum to join the perturbation process. 
Secondly, slight distortions of the octahedron result in anisotropic Yb-S orbital overlap. 
Both effects work together to select specific spatially anisotropic orbitals that participate in electron exchange processes. 

Our analysis yielded an average ratio of $\overline{J_{2}}/\overline{J_{1}}=0.9\sim 1.0$ 
and a typical ratio of $\overline{\Gamma_{\eta}}/\overline{J_{\eta}}=0.01\sim0.05$. 
However, the precise values may vary due to potential ambiguities in the lattice parameters of the material. 
The resulting Hamiltonian, expressed as Eq.(\ref{eq:ham}), 
is a simplification of the derived Hamiltonian XYZ+(off-diagonal) to the XXX+(xy+yx) model 
with $\Gamma_1/J_1=\Gamma_2/J_2$. 
Despite this simplification, we believe that our findings provide a sound explanation for the experimental observations in YbCuS$_2$.
\par
Based on our calculations with reference parameters $\phi_{\rm Yb}=0.23\pi$-$0.25\pi$ and $\theta_{\rm Yb}=0.003\pi$-$0.016\pi$ 
for YbCuS$_2$, and by comparing the phase diagram in Fig.\ref{f1}(a), 
we conclude that YbCuS$_2$ indeed hosts the ND phase, which comprehensively explains the experimental features reported thus far.
Firstly, our calculations at $\phi_{\rm Yb}$ and $\theta_{\rm Yb}$ reveal the structure factors of longitudinal 
$\langle S^z_i S^z_j\rangle$ and transverse $\langle S^x_i S^y_j\rangle$ correlations 
with an incommensurate wave number $q \sim 0.55\pi$-$0.6\pi$, which closely matches the 
experimentally reported wave number $k=(0.305,0.0)$ corresponding to $q=0.61\pi$ in the leg direction\cite{onimaru}. 
It may worth noting that such incommensurate diagonal $zz,xx$ correlations exists for $\Gamma=0$ \cite{Hikihara2001,Bursill1995} 
but the $xy$ ones are not and may have relevance to the rotating magnetic structure in the $xy$ plane. 
\par
Most importantly, our nematic-dimer (ND) state exhibits a robust spin-gapped singlet coexisting with 
a fluctuating nematic component, giving rise to the observed nonmagnetic gapless excitation that explains 
the $T$-linear behavior of $1/T_1$ in the $^{63/65}$Cu-NQR. 
Furthermore, this state transforms into a 1/3-plateau phase at a critical field $h\sim J_1/2$ in our phase diagram. 
In the experimental context, the average Curie-Weiss temperature is approximately 30 K, 
and the critical field of the plateau phase is 5 T, aligning well with our numerical predictions. 
\par
Finally, we emphasize the theoretically intriguing aspect of the ND phase in our phase diagram. 
In the standard nematic phase observed in spin ladders, 
the one-particle magnon excitation is gapped while the two-magnon excitation is gapless\cite{Hikihara2008,Lauchili2006}. 
It typically occurs due to the localization of magnons caused by frustration-induced hopping cancellation. 
In such cases, the two-magnon bound state tends to condense earlier than the individual magnons, 
particularly near the saturation field\cite{Zhitomirsky2010}. 
In our model, the spin gap or a quasi-one-magnon gap in the non-conserved $S^z$ 
indeed occurs by the localization of spin-1/2 singlet pairs due to the frustration effect, breaking the lattice symmetry. 
However, the corresponding one-magnon instability does not compete with the quasi-two-magnon instability, 
as the corresponding $S^z$-oriented magnetic order never occurs in our phase diagram. 
This fact allows the unusual coexistence of the spin gap and the nonmagnetic gapless excitation in a wide range of parameters in the phase diagram. 
As we have discussed in \S.\ref{sec:phaseboundary}, 
the ND phase has ${\mathbb Z}_2\otimes {\mathbb Z}_2$ but breaks the translational symmetry. 
The former breaks at the magnetic phase transition, and the latter further breaks or recovers, depending 
on which of the magnetic phases to enter. 
This kind of phenomena may have similarity with the layers of discussions on the several symmetry-breakings 
reported in the Heisenberg spin ladder with the Dzyaloshinskii-Moriya interaction\cite{Penc2007}.

\begin{acknowledgments}
We thank Takahiro Onimaru, Chikako Moriyoshi, Kenji Ishida, Shunsaku Kitagawa, 
Fumiya Hori, and Karlo Penc for discussions. 
This work is supported by the "The Natural Laws of Extreme Universe" (No. JP21H05191) KAKENHI for Transformative Areas 
from JSPS of Japan, and JSPS KAKENHI Grant No. JP21K03440. 
\end{acknowledgments}

\appendix

\section{Exact solutions}
\label{app:exact}
The phase diagram in Fig.~\ref{f1}(a) has multiple exact solutions: 
the MG line, MPS solution at the multicritical point\cite{Saito2024-2}, the nematic product state, 
the resonating valence bond state (RVB)\cite{Hamada1988}, 
the fully-polarized ferromagnetic state, and the product of the fully-polarized ferromagnetic states of the decomposed two chains. 
\par
{\it Majumdar-Ghosh(MG) line and multicritical point.} 
The $J_2/J_1=\Gamma_2/\Gamma_1=1/2$ is a MG line hosting singlet product state as an exact eigenstate, 
$|\Psi_{\rm MG} \rangle=\prod_{j=1}^{N/2} |s_j\rangle$ with 
$|s_j\rangle=(|\!\uparrow \downarrow\rangle-|\!\downarrow\uparrow\rangle)/\sqrt{2}$ on $[2j-1,2j]$ sites. 
The $\theta=0$ limit is the Majumdar-Ghosh model\cite{Majumdar1969,Majumdar1969-2}, 
and the upper endpoint with with $\Gamma_{\gamma}/J_{\gamma}=\sqrt{3}$ is the multicritical point 
$(\phi=0.1476\pi,\; \theta=\pi/3)$, with a ground state degeneracy of $(N+2)^2/4$ (even $N$), $(N+1)(N+3)/4$ (odd $N$) 
for an open boundary. The exact ground state can be obtained up to $N\sim100$, with $\sim3000$-fold degeneracy. 
The details of the method are given in Ref.[\onlinecite{Saito2024-3}]. 
\par
{\it Nematic product state. } The exact ground state is found at the isolated point $\;\Gamma_{\gamma}/J_{\gamma}=1$ and $J_2/J_1=-1/2$, 
$(\phi=0.852\pi,\;\theta=0.25\pi)$, at the center of the nematic (N) phase. 
Its form is given as $|\Psi\rangle=\prod_{i=1}^{N/2}|p_1\rangle_{2i-1,2i}$. 
Here, $|p_1\rangle=(|00\rangle+i|11\rangle)/\sqrt{2}$ is the eigenstate of the nematic order parameter 
$Q^{xy}=S^x_1S^y_2+S^y_1S^x_2$ with an eigenvalue $1/2$. 
\par
{\it $\Gamma=0$ exact solutions. } 
We briefly review the previously known exact ground state at $J_2/J_1=-1/4,\;\Gamma=0$ $(\phi=0.9220\pi,\;\theta=0)$. 
This point is at the boundary of the Haldane-dimer and ferromagnetic phases, 
and for PBC, the fully polarized $S_{{\rm tot}}=N/2$ state coexist with 
the nontrivial UDRVB state with $S_{{\rm tot}}=0$ which is analytically described by the equal weight superposition 
of all different choices of dimer covering states as 
$|\psi_{{\rm UDRVB}}\rangle = \sum_{i<j}\sum_{k<l}\cdots\sum_{m<n}[i,j][k,l]\cdots[m,n]$ 
where $[i,j]=(|01\rangle_{i,j}-|10\rangle_{i,j})/\sqrt{2}$. 
Along the line $0.9220\pi<\phi<1.5\pi$ with PBC the $S_{{\rm tot}}=N/2$ state continues to be the ground state. 
At the point $\phi=1.5\pi$ the zigzag chain is decomposed to the two ferromagnetic Heisenberg chains 
and the ground state is the product of two ferromagnetic chains. 
\par
Consider a unit triangle consisting of three sites, $[l-1,l,l+1]$, and define a Hamiltonian $h_l$ as
\begin{align} 
    h_l &= \frac{J_1}{2} (\bm{S}_{l-1}\cdot\bm{S}_{l}+\bm{S}_{l}\cdot\bm{S}_{l+1})
    +\frac{J_2}{2} \bm{S}_{l-1}\cdot\bm{S}_{l+1},
    \notag \\
    &+\frac{\Gamma_1}{2} (S^x_{l-1}S^y_{l}+S^y_{l-1}S^x_{l} + S^x_{l}S^y_{l+1}+S^y_{l}S^x_{l+1})
    \notag \\
    &+\frac{\Gamma_2}{2} (S^x_{l-1}S^y_{l+1}+S^y_{l-1}S^x_{l+1}).
    \label{eq:tr}
\end{align}
When diagonalizing $h_l$, we find that the lowest energy state has degeneracy of $D_g=6$ 
for the exact solution points and $D_g=4$ for the lines that connect the points. 
Along the solid lines the system has exact solutions and its subspace is the $D_g$ lowest energy state. 
The broken lines do not have the exact solutions. 
The relevance with Eq.(\ref{eq:tr}) and the exact solutions are explained in detail in Ref.[\onlinecite{Saito2024-3}], 
while we can explain here that the exact solution points are related and are not isolated. 

\section{Mean field approach around a multicritical point}
\label{app:mean}
\begin{figure}
    \centering
    \includegraphics[width=8cm]{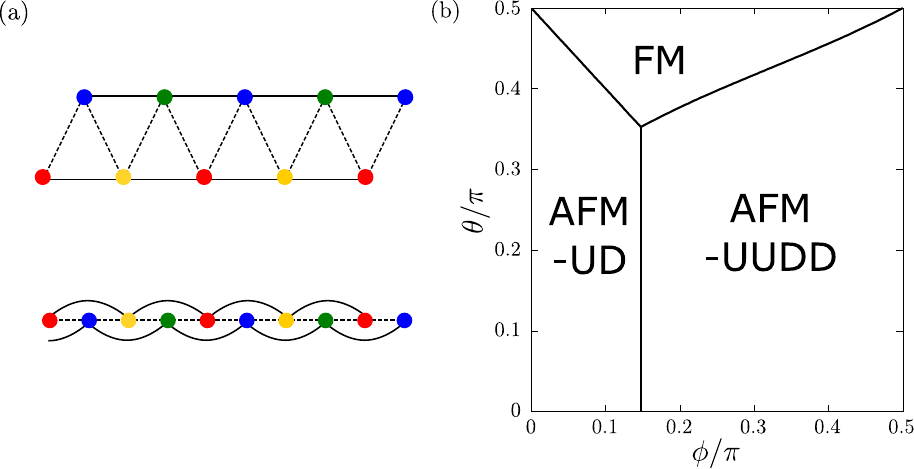}
    \caption{(a) Schematic illustration of coarse-graining a model with four sublattices. 
    (b) Mean-field phase diagram obtained by the competition of energies in Eq.(\ref{eq:E_r}). 
}
    \label{fap}
\end{figure}

At around the multicritical point in the phase diagram, 
the $J_1$-$J_2$-$\Gamma_1$-$\Gamma_2$ model described by Eq.(\ref{eq:ham}) suffers a competition among several orders. 
The magnetic orders to be considered here are ferromagnet (FM), two antiferromagnets with UD2 and UUDD configurations, 
and to accommodate all of them on equal footing, we need to consider the periodicity of magnetic moments up to four sites.  
The coordinate of the lattice sites are taken as $R=(4n-3)a,\cdots,4na$ with lattice constant $a$ 
where $n=1,\cdots, N/4$ is the unit cell index as shown in Fig.~\ref{fap}(a). 
We rely on the philosophy that the fluctuations around each order take place 
in a way that the magnetic moments belonging to the same magnetic sublattice are predominantly ferromagnetic. 
\par
We implement a scheme following Nelson and Fisher\cite{nelson1975} that takes a continuum limit, 
$\bm S_j\rightarrow \bm {S}(R)$, by retaining distinglishable four sublattices. 
The Hamiltonian in Eq.(1) in the main text is rewritten as 
\begin{align}
    {\mathcal H} &= \frac{1}{2} \sum_{R,R'} \{
        J(R-R')\bm{S}(R)\cdot \bm{S}(R') 
        \notag \\
        &+  \Gamma(R-R')(S^x(R) S^y(R') + S^y(R) S^x(R'))\}.
\end{align}
Then, we transform the spins as 
\begin{align}
&S^x = S^{x'}\cos(\pi/4)  + S^{y'}\sin(\pi/4) , \\
&S^y = -S^{x'}\sin(\pi/4)  + S^{y'}\cos(\pi/4), 
\end{align}
which is the rotation of $xy$-axes by $\pi/4$ about the $z$-axis where 
we adopt $x'y'z$-axes for spin coordinates in the following to distinguish from the original $xyz$-axes. 
Then, we find 
\begin{align}
    {\mathcal H}^{\pi/4} &= \frac{1}{2} \sum_{R,R'} \{
        J(R-R')\bm{S}(R)\cdot \bm{S}(R') 
        \notag \\
        & - \Gamma(R-R')(S^{x'}(R) S^{x'}(R') - S^{y'}(R) S^{y'}(R'))\}.
    \label{eq:Hpi4}
\end{align}
We now introduce the following unitary transformation about the spin operators inside the unit cell as 
\begin{equation}
\left(\begin{array}{l}
\bm S_{1,n} \\
\bm S_{2,n} \\
\bm S_{3,n} \\
\bm S_{4,n} 
\end{array}\right)
= \frac{1}{4}
\left(\begin{array}{rrrr}
1 & 1 & 1 & 1 \\
1 &-1 & 1 &-1 \\
1 & 1 &-1 &-1 \\
1 &-1 &-1 & 1 
\end{array}\right)
\left(\begin{array}{l}
\bm S((4n-3)a) \\
\bm S((4n-2)a) \\
\bm S((4n-1)a) \\
\bm S(4na) 
\end{array}\right). 
\label{eq:s1to4}
\end{equation}

The Fourier transformation for these new spin variables are 
\begin{equation}
\bm S_\nu(q)= \sum_{n=1}^{N/4} e^{iq 4na} \bm S_{\nu,n}, 
\label{eq:sqfourier}
\end{equation}
where $q$ runs over a folded Brillouin zone :
\begin{align}
    |q|\leq \frac{\pi}{4a}.
    \label{eq:q_range}
\end{align}
When $\langle \bm S_\nu(q) \rangle \ne 0$ we have magnetic orderings, 
where $\nu=1,2,3,$ and $4$ correspond to 
ferromagnet ($\nu=1$), UD antiferromagnet ($\nu=2$), and UUDD antiferromagnet ($\nu=3,4$). 
\par
Substituting Eqs.(\ref{eq:s1to4},\ref{eq:sqfourier}) to Eq.(\ref{eq:Hpi4}) we find 
\begin{align}
    {\cal H}^{\pi/4} = &\frac{4}{N}\sum_{i,j}\sum_q 
    \sum_{\alpha} \{r_{ij,\alpha=x',y',z}(q)S_{i\alpha}(q)S_{j\alpha}(-q)\},
    \label{eq:H^rot_int}
\end{align}
with couplings given as 
\begin{align}
    r_{11,x'}(q)&=(J_1-\Gamma_1)(3+\cos(4qa))\notag \\ &\quad +(J_2-\Gamma_2)(2+2\cos(4qa)),
    \notag \\
    r_{22,x'}(q)&=-(J_1-\Gamma_1)(3+\cos(4qa))\notag \\ &\quad +(J_2-\Gamma_2)(2+2\cos(4qa)),
    \notag \\
    r_{33,x'}(q)&=(J_1-\Gamma_1)(1-\cos(4qa))\notag \\ &\quad -(J_2-\Gamma_2)(2+2\cos(4qa)),
    \notag \\
    r_{44,x'}(q)&=-(J_1-\Gamma_1)(1-\cos(4qa))\notag \\ &\quad -(J_2-\Gamma_2)(2+2\cos(4qa)),
    \notag \\
    r_{12,x'}(q)&=-r_{21,x'}(q)=(J_1-\Gamma_1)i\sin(4qa),
    \notag \\
    r_{13,x'}(q)&=-r_{31,x'}(q)=(J_1-\Gamma_1)i\sin(4qa)\notag \\ &\quad +(J_2-\Gamma_2)2i\sin(4qa),
    \notag \\
    r_{14,x'}(q)&=r_{41,x'}(q)=-(J_1-\Gamma_1)(1-\cos(4qa)),
    \notag \\
    r_{23,x'}(q)&=r_{32,x'}(q)=(J_1-\Gamma_1)(1-\cos(4qa)),
    \notag \\
    r_{24,x'}(q)&=-r_{42,x'}(q)=(J_1-\Gamma_1)i\sin(4qa)\notag \\ &\quad +(J_2-\Gamma_2)2i\sin(4qa),
    \notag \\
    r_{34,x'}(q)&=-r_{43,x'}(q)=(J_1-\Gamma_1)i\sin(4qa). 
    \label{eq:rg}
\end{align}
The other couplings are related to those above; 
$r_{ij,y'}(q)$ are obtained from $r_{ij,x'}$ by taking $\Gamma_{\eta}\rightarrow-\Gamma_{\eta}$, 
and the $r_{ij,z}(q)$ are given by taking $\Gamma_{\eta}\rightarrow 0$. 
\par
At the multicritical point, we have $J_1=2J_2\equiv J$, $\Gamma_1=2\Gamma_2 = 2J$. 
The remarkable feature of Eq.(\ref{eq:H^rot_int}) is that different spin components do not couple, 
which means that the competitions among different Ising orders take place. 
They are exclusive and the ones that contribute to the lowest energy modes are 
$r_{11,x'}(q)$, $r_{22,y'}(q)$, $r_{33,y'}(q)$, and $r_{44,y'}(q)$. 
Therefore, we can only leave these modes around the critical point and 
taking the $q\sim 0$ modes, we find 
\begin{align}
{\cal H}^{\pi/4} \sim &\frac{1}{N} \Big(
r_{11,x'} S_{1x'} S_{1x'} + r_{22,y'}S_{2y'}S_{2y'}   + r_{33,y'} S_{3y'} S_{3y'} \Big)
\end{align}
where we dropped off $r_{44,y'}$ because the UDDU is equivalent to UUDD of $r_{33,y'}$, related by the translation. 
Here, we regard the spins as classical fields in the range $-N/4 \leq S_{i\alpha} \leq N/4$, 
and they take the minimum or maximum at $S_{i\alpha}=0, \pm N/4$. 
We thus have only three modes that contribute to the low energy excitation at the multicritical point. 
\par
The energies of the ordered state $i\alpha$ are given as 
$ E_{i\alpha} = r_{ii,\alpha}(0) N/16$, which are explicitly written as 
\begin{align}
 & E_{1x}=\frac{N}{4}(J_1-\Gamma_1+J_2-\Gamma_2), \quad\quad  (\text{FM}) \notag\\
 & E_{2y}=\frac{N}{4}(-J_1-\Gamma_1+J_2+\Gamma_2),\quad(\text{AFM-UD}) \notag\\ 
 & E_{3y}=\frac{N}{4}(-J_2-\Gamma_2). \rule{19mm}{0mm} (\text{AFM-UUDD})
    \label{eq:E_r}
\end{align} 
The competition of these three energies yields the phase diagram shown in Fig.~\ref{fap}(b). 
Combined with the Fig~\ref{f6}(b), it will explain the basic feature of the 
numerically accurate phase diagram in Fig.~\ref{f1}(a). 
\par
We note that Eq.(\ref{eq:E_r}) provides the same energies as 
the energy expectation values of the original Hamiltonian Eq.(\ref{eq:ham}) 
about the trial product wave functions of the corresponding 
FM1 phase, the AFM-UD2 phase the UUDD phase, e.g. 
$E_{1x}= \langle \Psi_{\rm FM}| {\cal H} |\Psi_{\rm FM}\rangle$ 
\begin{align}
&|\Psi_{\rm FM} \rangle= \prod_{j=1}^{N} |f\rangle_{j}, \quad 
 |\Psi_{\rm UD}\rangle= \prod_{j=1}^{N/2}|u\rangle_{2j-1} |d\rangle_{2j} ,\notag \\
& |\Psi_{\rm UUDD}\rangle= \prod_{j=1}^{N/4} |u\rangle_{4j-3} |u\rangle_{4j-2} |d\rangle_{4j-1} |d\rangle_{4j}, 
\label{eq:mfwf}
\end{align} 
where $|u\rangle_j=|\!\!\uparrow\rangle_j + e^{i\pi/4}|\!\!\downarrow\rangle_j$, $|d\rangle_j=|\!\!\uparrow\rangle_j - e^{i\pi/4} |\!\!\downarrow\rangle_j$, 
$|f\rangle_j=|\!\!\uparrow\rangle_j + e^{i3\pi/4}|\!\!\downarrow\rangle_j$ 
where the spin orientations are described in the $xyz$-frame and 
their magnetic moment points in the $(\pm1,\pm1,0)$ and $(\pm 1,\mp 1,0)$ directions, respectively. 

\section{Spin gap for various directions of magnetic field }
\label{app:spingap}
\begin{figure}
    \centering
    \includegraphics[width=8cm]{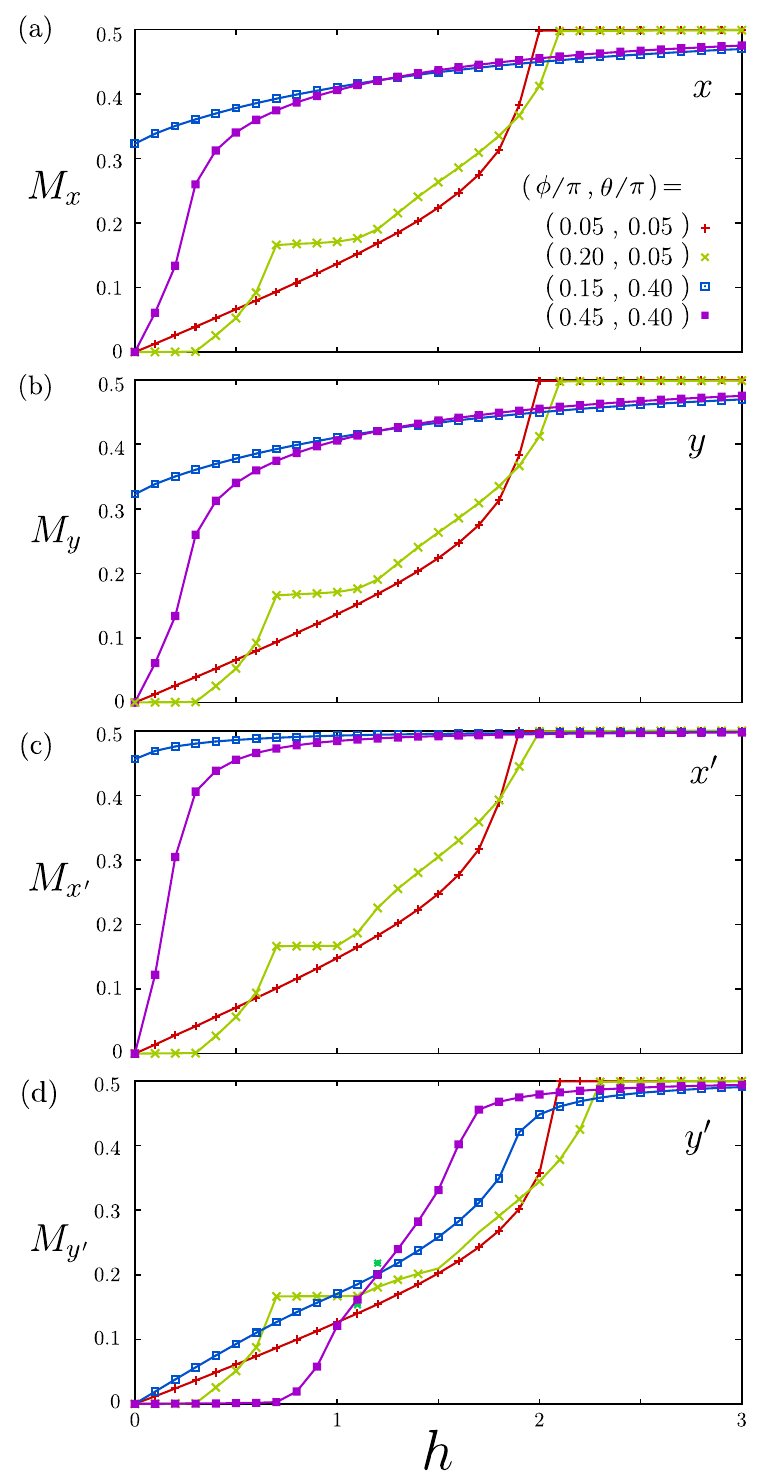} 
    \caption{ Magnetization curves for field $h$ applied along the 
    (a) $x$, (b) $y$, (c) $x'$, and (d) $y'$-directions. 
     We choose $(\phi,\theta)/\pi=(0.05,0.05)$ for TLL, $(0.2, 0.05)$ for ND, 
     $(0.15, 0.4)$ for FM1, and $(0.45, 0.4)$ for AFM-UUDD phases. 
}
    \label{fap2}
\end{figure}

Because the $z$-axis is a magnetic easy axis, the application of a magnetic field 
to the $z$-direction shown in Fig.~\ref{f5} may naturally open a spin gap. 
Whereas, the field applied inside the $xy$ plane may not necessarily be the case because 
the system does not magnetically order inside the $xy$ plane in the ND phase. 
We thus apply a field in the four different directions, $\alpha=x,y,x',$ and $y'$, 
where the latter two are the magnetic easy axes of the FM1,UD1 ($x'$) and FM2, UD2, UUDD($y'$). 
We add the Zeeman term in these directions, $-h\sum_i S_i^\alpha$, and measure $M_\alpha$ using SSD-DMRG. 
Figure~\ref{fap2} shows the magnetization curves for the four cases. 
We find that ND phase has a robust spin gap which does not depend much on the field direction, 
confirming the finite spin gap. 
We also show that TLL and UUDD phases are gapless. 
In the FM1 phase, the finite magnetization in the $x'$-direction from $h=0$ is observed. 

\section{Details of bond operator mean-field calculation}
\label{app:bop}
We present here the details of the bond-operator mean-field calculation in Section \ref{sec:bond}. 
\subsubsection{ Derivation of ${\cal H}_{\mathrm{bo}}$ } 
We first show the approximation we made in deriving the mean-field Hamiltonian ${\cal H}_{\mathrm{bo}}$. 
From Eq.(\ref{eq:bos_rep}), the commutation relations, and the constraint Eq.(\ref{eq:par_num}), we can rewrite the 
three inter-dimer spin-spin coupling terms as 
\begin{align}
    [S_1^{\alpha},S_1^{\beta}]&=i\epsilon_{\alpha \beta \gamma}S_1^{\gamma},\quad
    [S_2^{\alpha},S_2^{\beta}]=i\epsilon_{\alpha \beta \gamma}S_2^{\gamma}, \label{eq:S_com_1}\notag\\
    [S_1^{\alpha},S_2^{\beta}]&=0,
\end{align}
\begin{align}
    S_1^{\alpha}S_1^{\alpha}&=\frac{3}{4}
    +\frac{1}{4}(s^{\dagger}s^{\dagger}p_{\alpha}p_{\alpha}+ssp_{\alpha}^{\dagger}p_{\alpha}^{\dagger}\notag \\
    &-\sum_{\substack{\beta\neq\gamma\\ x,y,z}}p_{\beta}^{\dagger}p_{\beta}^{\dagger}p_{\gamma}p_{\gamma}) \notag \\
    &-\frac{1}{2}i\epsilon_{\alpha \beta \gamma}(s^{\dagger}p_{\alpha}p_{\beta}^{\dagger}p_{\gamma}
     + sp_{\alpha}^{\dagger}p_{\beta}^{\dagger}p_{\gamma}),
    \notag\\
    S_2^{\alpha}S_2^{\alpha}&=\frac{3}{4}
    +\frac{1}{4}(s^{\dagger}s^{\dagger}p_{\alpha}p_{\alpha}+ssp_{\alpha}^{\dagger}p_{\alpha}^{\dagger} \notag \\
    &-\sum_{\substack{\beta\neq\gamma\\ x,y,z}}p_{\beta}^{\dagger}p_{\beta}^{\dagger}p_{\gamma}p_{\gamma}) \notag \\
    &+\frac{1}{2}i\epsilon_{\alpha \beta \gamma}(s^{\dagger}p_{\alpha}p_{\beta}^{\dagger}p_{\gamma} 
     + sp_{\alpha}^{\dagger}p_{\beta}^{\dagger}p_{\gamma}),
    \notag\\
    S_1^{\alpha}S_2^{\alpha}&=-\frac{3}{4}s^{\dagger}s
      + \frac{1}{4}p_{\alpha}^{\dagger}p_{\alpha}
    -\frac{1}{4}(s^{\dagger}s^{\dagger}p_{\alpha}p_{\alpha}  \notag \\ 
    & +ssp_{\alpha}^{\dagger}p_{\alpha}^{\dagger} \notag \\
    &+\sum_{\substack{\beta\neq\gamma\\ x,y,z}}p_{\beta}^{\dagger}p_{\beta}^{\dagger}p_{\gamma}p_{\gamma}) ,
    \label{eq:S_com}
\end{align}
\subsubsection{Diagonalizing Eq.(\ref{eq:H_bo_compact}) }
We explicitly show the form of the matrices consisting Eq.(\ref{eq:H_bo_compact}) as 
\begin{align}
    A(k) = 
    \begin{pmatrix}
        a(k) & c(k) & b(k) & d(k) \\
        c(k) & a(k) & d(k) & b(k) \\
        b(k) & d(k) & a(k) & c(k) \\
        d(k) & b(k) & c(k) & a(k)
    \end{pmatrix}
    \label{eq:AB_1}\\
    B(k) =  
    \begin{pmatrix}
        a(k) & b(k) \\
        b(k) & a(k)
    \end{pmatrix} ,
    \label{eq:AB_2}
\end{align}
\begin{align}
    a(k) &= \frac{J_1-\mu}{2} + \lambda\frac{-J_1+2J_2}{4}\bar{s}^2 \cos k\hat{z},
    \label{eq:abcd_1}\\
    b(k) &= \lambda\frac{-J_1+2J_2}{4}\bar{s}^2 \cos k\hat{z},
    \label{eq:abcd_2}\\
    c(k) &= -\frac{J_1}{4}\gamma + \lambda \frac{-J_1+2J_2}{4}\gamma\bar{s}^2 \cos k\hat{z},
    \label{eq:abcd_3}\\
    d(k) &= \lambda \frac{-J_1+2J_2}{4}\gamma\bar{s}^2 \cos k\hat{z},
    \label{eq:abcd_4}
\end{align}

Using a $4\times 4$ real matrix $L_k$ and a $2\times 2$ real matrix $M_k$, we can diagonalize A(k) by the Bogoliubov transformation:
\begin{align}
    L_k \bm{u}_k = \bm{u}_k ',
    \notag \\
    \bm{u}_k' = (\tilde{p}_{k,1} ,\tilde{p}_{k,2},\tilde{p}_{-k,1}^{\dagger} ,\tilde{p}_{-k,2}^{\dagger})^T,
    \label{eq:Bog_1}\\
    M_k \bm{v}_k = \bm{v}_k ',
    \notag \\
    \bm{v}_k ' = (\tilde{p}_{k,3}, \tilde{p}_{-k,3}^{\dagger})^T. 
    \label{eq:Bog_2}
\end{align}
Then, we find that 
\begin{align}
    &\sum_k \bm{u}_k^{\dagger} A(k) \bm{u}_k 
    \notag \\
	    &= \sum_k \left( \omega_1(k)\tilde{p}_{k,1}^{\dagger}
	    \tilde{p}_{k,1} + \omega_2(k)\tilde{p}_{k,2}^{\dagger}
	    \tilde{p}_{k,2} + \frac{\omega_1(k)+\omega_2(k)}{2}\right),
    \label{eq:xy_diag}\\
    \omega_{1} &= 2\sqrt{(a+c)^2-(b+d)^2},
    \label{eq:xy_diag_2} \\
    \omega_{2} &= 2\sqrt{(a-c)^2-(b-d)^2},
    \label{eq:xy_diag_3}
\end{align}
and 
\begin{align}
    \sum_k \bm{v}_k^{\dagger}B(k)\bm{v}_k
    &= \sum_k \left(\omega_{3}(k)\tilde{p}_{k,3}^{\dagger}
	    \tilde{p}_{k,3}+\frac{\omega_{3}(k)}{2}\right),
    \label{eq:quadratic_diag_1}\\
    \omega_{3} &= 2\sqrt{a^2-b^2}.
    \label{eq:quadratic_diag_2}
\end{align}
The ground state energy is given as 
\begin{align}
   & E_{\mathrm{GS}}^{\mathrm{MF}} 
    = \sum_k \Big( \frac{5}{2}\mu - \frac{3}{2}J_1 - \mu\bar{s}^2  
       + \frac{\omega_1(k) + \omega_2(k) + \omega_3(k)}{2} \Big)
      \notag \\
   &\rule{8mm}{0mm} = N_{\mathrm{d}} \Big( \frac{5}{2}\mu - \frac{3}{2}J_1 - \mu\bar{s}^2 
      + \sum_{l=1}^3 C_l\; E(X_l) \Big), 
      \notag \\
   & C_1= \frac{1}{\pi} \Big( |\epsilon_{p_1}\!-\mu| 
       \big( |\epsilon_{p_1}\!-\mu|+|\lambda(1-\gamma)(-J_1+2J_2)\bar{s}^2|\big)\Big)^{1/2},
    \notag \\ 
   & X_1= \Big(\frac{2|\lambda (1-\gamma)(-J_1+2J_2)\bar{s}^2|}{|\epsilon_{p_1}-\mu| +|\lambda (1-\gamma)(-J_1+2J_2)\bar{s}^2|}\Big)^{1/2},
     \notag \\  
   & C_2 = \frac{1}{\pi} \Big( |\epsilon_{p_{-1}}\!-\mu| 
       \big( |\epsilon_{p_{-1}}\!-\mu|+|\lambda(1+\gamma)(-J_1+2J_2)\bar{s}^2|\big)\Big)^{1/2},
      \notag \\ 
   & X_2= \Big(\frac{2|\lambda (1+\gamma)(-J_1+2J_2)\bar{s}^2|}{ |\epsilon_{p_{-1}}-\mu| + |\lambda (1+\gamma)(-J_1+2J_2)\bar{s}^2|}\Big)^{1/2},
      \notag \\ 
   & C_3 =\frac{1}{\pi}\Big( |\epsilon_{p_{0}}\!-\mu| (|\epsilon_{p_{0}}\!-\mu|+|\lambda(-J_1+2J_2)\bar{s}^2|\Big)^{1/2},
      \notag \\ 
   & X_3 = \Big( \frac{2|\lambda (-J_1+2J_2)\bar{s}^2|}{|\epsilon_{p_{0}}-\mu| + |\lambda (-J_1+2J_2)\bar{s}^2|}\Big)^{1/2}, 
    \label{eq:E_GS} 
\end{align}
where we use the complete elliptic integral of the second kind with $0\leq m\leq 1$, 
\begin{align}
    E(m) = \int_0^{\frac{\pi}{2}}\sqrt{1-m^2 \sin^2 \theta} \ d\theta,
    \label{eq:comp_elli}
\end{align}
We searched for the solutions $(\mu,\bar{s}^2)$ by numerically 
over the parameter space that gives $\omega_l \in {\mathbb R}$ 
for $\min_{\mu,\bar{s}^2} E_{\mathrm{GS}}^{\mathrm{MF}}$: 
\begin{align}
    & |\epsilon_{p_{1}}-\mu| \geq \left|\lambda(1-\gamma)(-J_1+2J_2)\bar{s}^2\right|,
    \notag\\
    & |\epsilon_{p_{-1}}-\mu| \geq \left|\lambda(1+\gamma)(-J_1+2J_2)\bar{s}^2\right|,
    \notag\\
    & |\epsilon_{p_{0}}-\mu| \geq \left|\lambda(-J_1+2J_2)\bar{s}^2\right|,
    \notag\\
    &0 \leq \bar{s}^2 \leq 1,
    \notag\\
    &0 \leq \mu \leq 1.
    \label{eq:ineq}
\end{align} 


\bibliography{zigzag_nematic_ref}
\bibliographystyle{apsrev4-1}

\end{document}